\documentclass{jfm}

\usepackage{graphicx}
\usepackage{newtxtext}
\usepackage{newtxmath}
\usepackage{caption}
\usepackage{slashed}
\usepackage{subfigure}
\usepackage{overpic}
\usepackage{soul}
\usepackage{natbib}
\usepackage{bm}
\usepackage{amsmath,amssymb,mathtools}
\usepackage{hyperref}
\usepackage{adjustbox}
\hypersetup{
	colorlinks = true,
	urlcolor   = blue,
	citecolor  = blue,
	linkcolor  = blue,
}

\newcommand{\RomanNumeralCaps}[1]

\title{Causal structures of turbulent skin-friction drag in wall-bounded turbulent flows}

\author{Yunchao Zhao\aff{1}, Yitong Fan\aff{1}, \and Weipeng Li\aff{1,}\corresp{\email{liweipeng@sjtu.edu.cn}}}

\affiliation{\aff{1}School of Aeronautics and Astronautics, Shanghai Jiao Tong University, Shanghai 200240, China}
\begin{document}
	\maketitle
	\begin{abstract}
		Understanding the mechanism of turbulent skin-friction drag (TSD) generation is of fundamental and practical importance for designing effective drag reduction strategies.
		However, many previous studies adopted correlation analysis to reveal the causal map between turbulent motions and TSD generation, an approach that is potentially risky as correlation does not necessarily imply causation.
		In this study, a novel causal inference method called Liang-Kleeman information flow (LKIF) is utilized for the first time to identify the velocity-induced causal structures related to TSD generation in a turbulent channel flow.
		The statistical properties of the causal structures are comprehensively investigated.
		The positive and negative causal structures, defined by their signs and respectively associated with an increase and decrease in TSD information entropy, promote and suppress the generation of extreme TSD.
		Particularly, we find that the underlying physics of causal structures is essentially associated with the processes of streamwise streaks and rolls approaching or receding from the extreme events.
		Results indicate that the physics-informed LKIF framework can reveal a more explicit and interpretable causal relationship than correlation analysis.
	\end{abstract}
	
	
	\section{Introduction}\label{sec1}
	
	Any object moving through a fluid suffers a skin-friction drag due to the no-slip condition at the wall surface.
	The skin-friction drag produced in a wall-bounded turbulent flow has been identified to be much higher than that in a laminar case, which directly impacts the aerodynamic/hydrodynamic performance of aircraft, ships, pipelines, and wind turbines, to name a few.
	For instance, the turbulent skin-friction drag (TSD) accounts for up to 50\% of the total drag in commercial aircraft, 90\% in submarines, and nearly 100\% in long pipes and channels~\citep{Marusic2010,Li2019}.
	A long-standing challenge has been to understand the causal relationship of TSD generation and its associated near-wall dynamics, which are of fundamental and practical importance for designing effective drag reduction strategies that can lead to ecological and economic benefits by lowering energy consumption and operational costs~\citep{Marusic2021}.
	
	It is widely accepted that the TSD generation is closely linked to the dynamical motions of well-organized coherent structures in the turbulent boundary layers, including streaks and quasi-streamwise vortices in the near-wall region~\citep{Kravchenko1993,Choi1994,Jeong1997} and large-/very-large-scale motions populating in the logarithmic and outer regions~\citep{Abe2004,Marusic2010a}.
	The formation and breakdown of streaks dominate the near-wall quasi-periodic self-sustaining process~\citep{Hwang2015,Hwang2016}, which is regarded as the direct causality contributing to the TSD generation, particularly through the `bursting' events associated with energetic sweeps that transport high-momentum fluids towards the wall~\citep{Kim1971,Flores2010,Jimenez2025}.
	Meanwhile, the outer large-scale motions influence the wall shear stresses in terms of superposition and modulation on the near-wall small-scale structures via nonlinear interactions~\citep{Marusic2010,Hutchins2007,Mathis2009,Mathis2011,Agostini2019}.
	The aforementioned coherent structures and their relationships with TSD generation are mostly identified with correlation analysis~\citep{Wallace2014}.
	However, using correlation to infer causation is risky, as correlation does not necessarily imply causation, whereas causation definitely includes correlation.
	It happens because correlation does not carry the required directionality and asymmetry to infer causation~\citep{Kantz2003,Beebee2009,Pearl2009,Liang2014,Lozano-Duran2020,Wang2021}.
	Therefore, the underlying causal relationships between turbulent motions and TSD generation have not been well-established based on the conclusions drawn from the correlation analysis.
	
	The identification of causality in complex systems (such as wall-bounded turbulent flows) presents significant challenges and remains an open question.
	Granger Causality (GC) offers a framework that uses predictability instead of correlation to infer causal relationships~\citep{Granger1969}.
	Specifically, the predictability means that, GC detects the causal map of variable $X_2\to X_1$ by evaluating the reduction in the prediction error of $X_1$ achieved when incorporating the historical information of $X_2$ into the set of potential causative variables.
	Crucially, Granger's framework fundamentally requires system separability (i.e. a causative factor is exclusive to only one variable) which cannot be guaranteed in the presence of latent confounding variables.
	For non-separable systems, Convergent Cross Mapping (CCM) has been proposed as an effective methodology to discover the causal map of $X_2\to X_1$ by measuring the capability to reconstruct the states of $X_2$ via the past state of $X_1$~\citep{George2012}.
	Moreover, CCM demonstrates superior performance in handling systems characterized by weak-to-moderate coupling strengths.
	Despite their utility, the metrics of GC and CCM are not grounded in physical interaction with causality, thus failing to possess a true causal attribute.
	To address this limitation, information-theoretic metrics have gained prominence in causal inference.
	Notably, Transfer Entropy (TE)~\citep{Schreiber2000} and its variants~\citep{Verdes2005,Staniek2008,Pompe2011,Kugiumtzis2013,Lozano-Duran2022} quantify the causal map of $X_2\to X_1$ by measuring the reduction of Shannon entropy when forecasting the future state of $X_1$ upon incorporating the historical information of $X_2$.
	These approaches facilitate the quantification of causal influence strength among variables.
	Recently, artificial intelligence techniques have been increasingly employed for causal discovery (e.g.~\citealp{Hill2019} and \citealp{Lagemann2023}) due to their potential in addressing some conditions that are often intractable for classic approaches, such as high dimensionality and data limitation.
	
	Based on information theory, an alternative approach, referred to as `Liang-Kleeman Information Flow' (LKIF), was initially developed for causality identification in simple bivariate dynamical systems~\citep{Liang2005}, and has been further extended to handle more complex systems~\citep{Liang2014,Liang2008,Liang2015,Liang2016,Liang2021}.
	Analogous to the concept of TE, the causality under the framework of LKIF is able to quantify the rate of information flow from one time series to another.
	The key difference between LKIF and TE is that LKIF is derived from the governing equations of a deterministic or stochastic dynamical system followed by both cause and effect series, whereas TE does not rely on the explicit knowledge of governing equations, but only on the Shannon entropy and conditional Shannon entropy of variables themselves.
	From this perspective, LKIF appears to be fairly physics-informed.
	LKIF formalism has attracted a great deal of attention and has been applied in diverse scientific disciplines, including marine science~\citep{Liang2014,Li2024,Li2025}, economic science~\citep{Liang2015}, atmospheric science~\citep{Stips2016}, glaciology~\citep{Vannitsem2019}, and neuroscience problems~\citep{Hristopulos2019}, to name a few.
	
	Causal inference has been incrementally used in turbulence research.
	GC was first adopted to study the energy redistribution mechanisms between elongated scales in the logarithmic layer~\citep{Tissot2014}.
	The capability of TE was highlighted by the studies focusing on, such as the cascading process~\citep{Materassi2014}, the energy-containing eddies~\citep{Lozano-Duran2020}, surface–subsurface interactions between free turbulent flow and porous medium~\citep{Wang2021}, reduced-order modeling and turbulent controlling~\citep{Lozano-Duran2022}, large-scale coherent structures~\citep{Martinez-Sanchez2023}, and developing both data-driven error-controlling schemes and wall models~\citep{Massaro2023}.
	Nevertheless, TE is criticized due to its requirement of long time series and prohibitive cost when computing probability density functions, especially in high dimensional cases~\citep{Liang2014,Hlavackova-Schindler2007,Runge2012}.
	Apart from aforementioned applications of classic methods, some studies also adopt other causality discriminance, such as Perron–Frobenius operator~\citep{Jimenez2023} and performing interventional numerical experiment~\citep{Lozano-Duran2021}.
	The interventional framework, which does not rely on the information theory, might face the potential drawbacks of the high cost of repetitive computations and possible omission of relevant causal factors~\citep{Jimenez2025}.
	With respect to LKIF, it seems to be overlooked in the studies of turbulent flows, except for \cite{Liang2016a} and \cite{Zhang2024}.
	\cite{Liang2016a} inspected the causal structure between the streamwise rolls and streaks, and \cite{Zhang2024} unveiled causality between self-sustained near-wall cycle and outer motions.
	In particular, \cite{Zhang2024} highlighted that both LKIF and TE can capture the major self-sustaining mechanism of near-wall turbulence, whereas LKIF runs 100 times faster than TE with respect to the computational efficiency.
	So far, the features of causal structures linked to TSD generation have not yet been extensively explored, which motivates this causality study via LKIF.
	
	The remaining of the paper is organized as follows.
	A brief introduction to LKIF formalism is given in § \ref{sec2}.
	Details of the direct numerical simulation (DNS) database and LKIF computation framework in the database are described in § \ref{sec3}.
	In § \ref{sec4}, statistical properties of the causal structures associated with TSD generation are investigated in detail, including distribution characteristics in § \ref{sec4.1}, comparisons with correlation analysis as well as advection velocity in § \ref{sec4.2}, relations with streaks/rolls in § \ref{sec4.3}, and their quadrant decomposition in § \ref{sec4.4}.
	Finally, concluding remarks are summarized in § \ref{sec5}.
	
	\section{Methodology of LKIF}\label{sec2}
	
	The Liang-Kleeman information flow (LKIF) got its name from the study of~\cite{Liang2005}, which was established on bivariate (two-dimensional) deterministic dynamical systems.
	It was further generalized to multivariate (high-dimensional) systems~\citep{Liang2007a} and stochastic systems~\citep{Liang2008}.
	In this study, we adopt the bivariate LKIF, and introduce the theory of LKIF briefly here.	
	Consider that two time series $X_1$ and $X_2$ are state variables in a stochastic dynamical system:
	\begin{equation}\label{eqB1}
		\mathrm{d}\bm{X}=\bm{F}\left(\bm{X},t\right)\mathrm{d}t+\bm{B}\left(\bm{X},t\right)\mathrm{d}\bm{W},
	\end{equation}
	where $\bm X=\left(X_1, X_2\right)^\mathrm{T}$ is the vector of state variables, $\bm F=\left(F_1, F_2\right)^\mathrm{T}$ the vector of differentiable functions modeling the deterministic part of the system, $t$ the time, $\bm{B}=\left(b_{ij}\right)$ a $2\times2$ matrix of diffusion coefficients characterizing the stochastic term, and $\bm W=\left(W_1, W_2\right)^\mathrm{T}$ a vector of independent standard Wiener processes.
	The information carried by the continuous variable $X_1$ is defined by the differential entropy:
	\begin{equation}\label{eqB2}
		H_1\coloneqq-\int_{\mathbb{R}}\rho_1\log_2\rho_1\mathrm{d}x_1,
	\end{equation}
	where $\rho_1$ is the marginal probability density function of $X_1$, and $x_1$ the value of the state variable $X_1$.
	A variable causal to $X_1$ intervenes in the time evolution of $H_1$ and therefore, presumably affects its time derivative $\mathrm{d}H_1/\mathrm{d}t$.
	It is decomposed into three terms as:
	\begin{equation}\label{eqB3}
		\frac{\mathrm{d}H_1}{\mathrm{d}t}=\frac{\mathrm{d}H_1^*}{\mathrm{d}t}+\frac{\mathrm{d}H_1^\mathrm{noise}}{\mathrm{d}t}+T_{2\to1},
	\end{equation}
	where $\mathrm{d}H_1^*/\mathrm{d}t$ represents contribution from $X_1$ itself, $\mathrm{d}H_1^\mathrm{noise}/\mathrm{d}t$ from stochastic noise arisen from the Wiener process, and $T_{2\to1}$ from $X_2$.
	The causality of $X_2$ to $X_1$ is quantified by $T_{2\to1}$, namely, the aforementioned Liang-Kleeman information flow.
	Based on the system~\eqref{eqB1}, $T_{2\to1}$ is given by:
	\begin{equation}\label{eqB4}
		T_{2\to1}=-\mathbb{E}\left[\frac{1}{\rho_1}\frac{\partial\left(F_1\rho_1\right)}{\partial x_1}\right]+\frac{1}{2}\mathbb{E}\left[\frac{1}{\rho_1}\frac{\partial^2\left(g_{11}\rho_1\right)}{\partial x_1^2}\right],
	\end{equation}
	where $\mathbb{E}$ represents the mathematical expectation with respect to $\bm X$, and $g_{11}=b_{11}^2+b_{12}^2$.
	A non-zero $T_{2\to1}$ means that $X_2$ is causal to $X_1$, and further, a positive/negative $T_{2\to1}$ means that $X_2$ makes $X_1$ more uncertain/certain.
	If $T_{2\to1}=0$, $X_2$ is not causal to $X_1$.
	A greater value of $|T_{2\to1}|$ means a stronger causality.
	The information flow in the opposite direction, $T_{1\to2}$, can be directly written out by interchanging the indices 1 and 2.
	
	In spite of the explicit expression of $T_{2\to1}$, Equation~\eqref{eqB4} is still hard to be estimated.
	Under the assumptions of linearity and Gaussianity of the probability distribution of the variables, Equation~\eqref{eqB4} can be approximated by Maximum Likelihood Estimation (MLE)~\citep{Liang2014} as follow:
	\begin{equation}\label{eqB5}
		\hat{T}_{2\to1}=\frac{C_{11}C_{2,\mathrm{d}1}-C_{12}C_{1,\mathrm{d}1}}{C_{11}C_{22}-C_{12}^2}\frac{C_{12}}{C_{11}},
	\end{equation}
	where $C_{ij}$ is the sample covariance between $X_i$ and $X_j$ ($i,j\in\{1,2,\mathrm{d}1,\mathrm{d}2\}$), $X_{\mathrm{d}i}$ the time-difference series of $X_i$.
	Both Equations~\eqref{eqB4} and~\eqref{eqB5} suggest the asymmetry of the causality, i.e.~$T_{2\to1}\neq T_{1\to2}$, which is a vital property that the correlation does not involve.
	The concise form of Equation~\eqref{eqB5} brings convenience to quantify causality since it only involves simple algebraic operations of the sample covariances.
	Hereafter, we will ignore the hat notation of estimator $\hat{T}_{2\to1}$ for brevity.
	To assess the statistical significance of $T_{2\to1}$, a significance test called Wald test is performed (see Appendix~\ref{appA}).
	Despite the assumption of linear evolution, the feasibility and power of~\eqref{eqB5} applying to complicated nonlinear systems have also been validated~\citep{Stips2016,Vannitsem2019,Hristopulos2019,Li2024,Li2025}.
	
	It should be noted that although Equation~\eqref{eqB5} is derived from the stochastic dynamical system~\eqref{eqB1}, it is also applicable for a deterministic one whose governing equation has the same form as \eqref{eqB1} with $\bm{B}=0$.
	Moreover, although the computation of $T_{2\to1}$ seems to depend solely on the data itself instead of the governing equations, which is identical to the drawback of TE we have claimed in the Introduction, we should clarify that this apparent dependence is just in form.
	The governing equation has been embedded in the whole derivation, and therefore, the Equation~\eqref{eqB5} itself is physics-informed.
	It is not computing the covariances but how to combine them that depends on the governing equations.

	\section{DNS database and LKIF computational framework}\label{sec3}
	
	\subsection{DNS database}\label{sec3.1}
	
	The DNS database of a turbulent channel at $Re_\tau\approx183$ is computed with a pseudospectral code~\citep{Kim1987}, where $Re_\tau$ is friction Reynolds number.
	Fourier discretization is used in the streamwise and spanwise directions, and Chebyshev polynomials are used in the wall-normal direction.
	The time marching scheme is a third-order Runge-Kutta scheme for nonlinear terms and an implicit Euler scheme for viscous terms.
	Periodic boundary conditions are used in the homogeneous directions, and no-slip boundary conditions are employed on the wall surfaces.
	The pressure gradient drives the flow to keep the mass flow constant.
	The flow fields are stored with a time interval of $\Delta t_\mathrm{s}^+=1.002$, to meet the demand of highly time-resolved time series.
	Details of parameters information are listed in table~\ref{tab1}.
	
	\begin{table}
		\begin{center}
			\def~{\hphantom{0}}
			\begin{tabular}{cccccccccccc}
				$\Rey_\tau$ & $N_x$ & $N_z$ & $N_y$ & $N_t$ & $L_x/h$ & $L_z/h$ & $L_y/h$ & $\Delta y_{\min}^+$ & $\Delta y_{\max}^+$ & $\Delta t_\mathrm{s}^+$ & $Tu_\tau/h$\\[3pt]
				183 & 256 & 257 & 129 & 8000 & 4$\pi$ & 2$\pi$ & 2 & 0.0552 & 4.50 & 1.002 & 43.73\\
			\end{tabular}
			\caption{Simulation parameters of the DNS database.
				$N_x$, $N_z$ and $N_y$ are grid numbers in the streamwise, spanwise and wall-normal direction, respectively.
				$N_t$ is the number of collected snapshots used for statistics.
				$L_x$, $L_z$ and $L_y$ are sizes of the computation domain in three directions.
				$h=1$ is the channel half-height.
				The $\Delta y_{\min}^+$ and $\Delta y_{\max}^+$ are the finest and coarsest grid spacings in wall-normal direction.
				$\Delta t_\mathrm{s}^+$ is the time interval between snapshots.
				$Tu_\tau/h$ is the total time of collected snapshots.}
			\label{tab1}
		\end{center}
	\end{table}

	\subsection{LKIF computational framework}\label{sec3.2}
	
	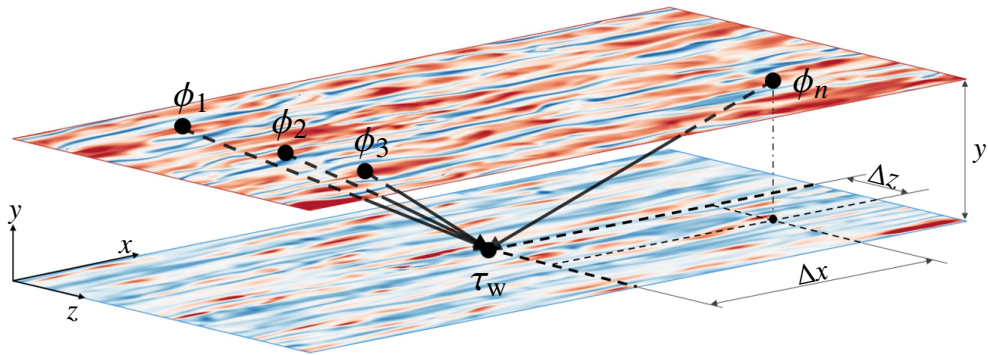
\begin{figure}
		\centering
		\includegraphics[width=5in]{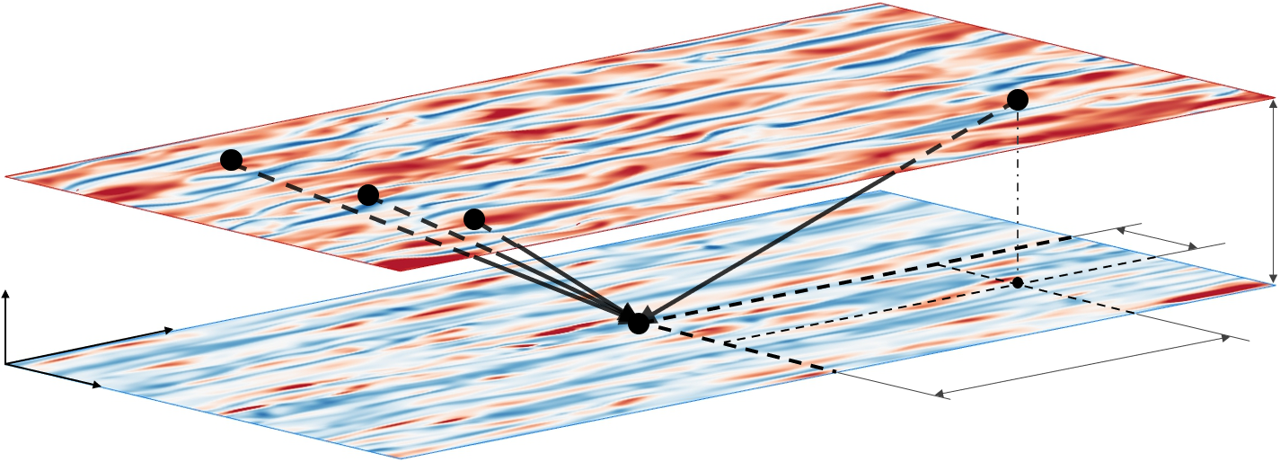}
		\put(1,74){\large$y$}
		\put(-66,26){\large\rotatebox{12}{$\Delta x$}}
		\put(-39,65){\large\rotatebox{-14}{$\Delta z$}}
		\put(-188,24){\Large$\tau_\mathrm{w}$}
		\put(-300,93){\Large$\phi_1$}
		\put(-263,83){\Large$\phi_2$}
		\put(-232,76){\Large$\phi_3$}
		\put(-67,98){\Large$\phi_n$}
		\put(-362,51){\large$y$}
		\put(-320,38){\large$x$}
		\put(-340,13){\large$z$}
		\caption{Schematic for computing LKIF ($T_{\phi\to\tau_\mathrm{w}}$) between the flow quantity $\phi$ and wall-shear stress $\tau_\mathrm{w}$.}
		\label{fig1} 
	\end{figure}

	To ascertain how a fluid event (characterized by a specific flow quantity $\phi$) acts on TSD generation (characterized by wall-shear stress $\tau_\mathrm{w}=\mu\partial u/\partial y|_{y=0}$, where $\mu$ is dynamic viscosity), figure~\ref{fig1} illustrates the schematic for computing LKIF ($T_{\phi\to\tau_\mathrm{w}}$), where $\tau_\mathrm{w}$ serves as the effect time series located at the center of the wall surface ($y=0$), and $\phi$ represents the causal time series assigned at a position with a distance ($\Delta x$, $y$, $\Delta z$) from the target $\tau_\mathrm{w}$.
	Hereafter, we use $x$, $y$ and $z$ to denote the streamwise, wall-normal and spanwise directions, and employ $u'$, $v'$ and $w'$ to denote the corresponding velocity fluctuations.	
	The flow quantity $\phi$ is individually specified as $u'$, $v'$ or $w'$, which may properly characterize the importance of streamwise streaks and bursting processes~\citep{Ling2024}.
	Moreover, setting a time lag of $\Delta t$ on the causal $\phi$ results in spatio-temporal variations of $T_{\phi\to\tau_\mathrm{w}}\left(\Delta x,y,\Delta z,\Delta t\right)$, helping us to analyze the historical causal relationship between $\phi$ and $\tau_\mathrm{w}$.
	Details about the database and calculation of $T_{\phi\to\tau_\mathrm{w}}\left(\Delta x,y,\Delta z,\Delta t\right)$ can be found in Appendix~\ref{appB}.
	Notably, the statistical significance of the following results has been rigorously examined with significance tests at a level of $\alpha=0.01$, providing 99\% confidence that the structures identified through LKIF are reliable (see Appendix~\ref{appD}).
	Additionally, we also provide the results at a higher friction Reynolds number of $Re_\tau\approx548$ in Appendix~\ref{appE}, and the results suggest a certain degree of Reynolds-number-invariance.
	
	\section {Results and discussion}\label{sec4}
	
	\subsection{Overview of the velocity-induced causal structures}\label{sec4.1}
	
	\begin{figure}
		\centering
		\subfigure{
			\begin{overpic}[width=2.3in]{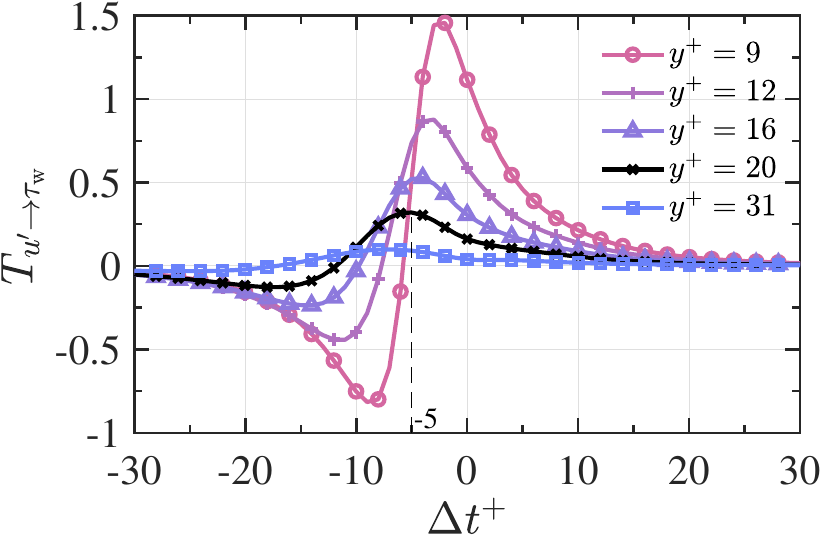}
				\put(1,66){{($a$)}}
			\end{overpic}\label{fig2a}}
		\subfigure{
			\adjustbox{valign=t, raise=4.32cm}{
				\begin{overpic}[width=2.59in]{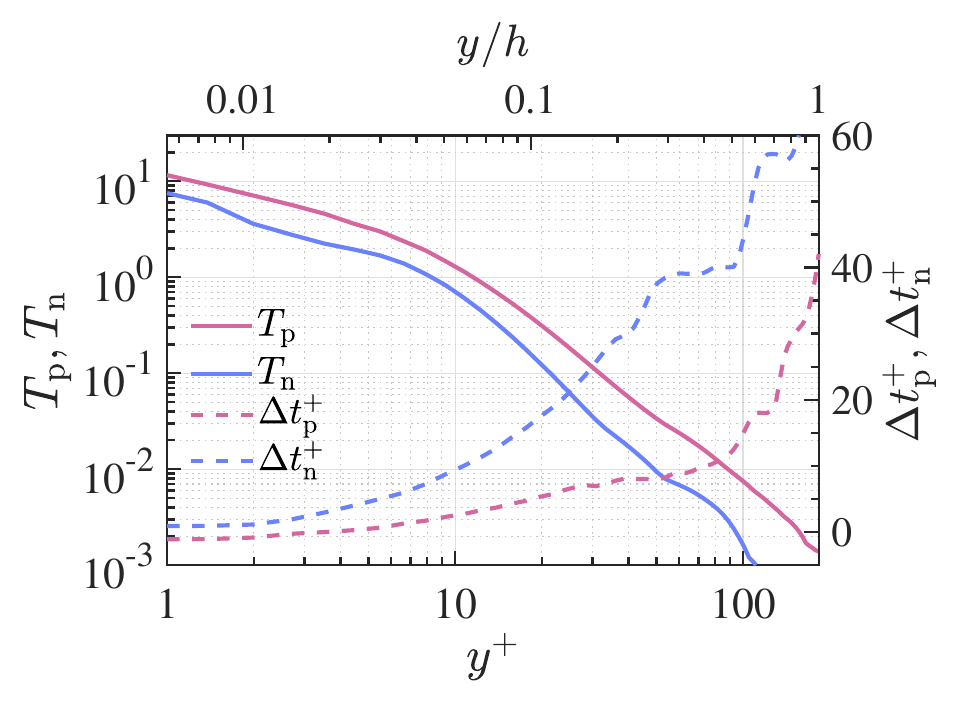}
					\put(1,63){($b$)}
				\end{overpic}\label{fig2b}}}
		\caption{LKIF from $u'$ at five specific heights to the target $\tau_\mathrm{w}$.
			($a$) Profiles of $T_{u'\to\tau_\mathrm{w}}$ versus $\Delta t^+$.
			($b$) Profiles of the positive and negative peak values ($T_{\rm p}$ and $T_{\rm n}$) versus $y$, as well as their corresponding time lags.}
		\label{fig2}
	\end{figure}

	We first allocate the causal signal on the streamwise velocity fluctuation $u'$, considering its dominant role in the momentum transport and energy production in the near-wall region, which are closely related to the TSD generation.
	Figure~\ref{fig2a} shows the variations of $T_{u'\to\tau_\mathrm{w}}$ as a function of $\Delta t^+$, where $u'$ locates right above the target $\tau_\mathrm{w}$, i.e. $\Delta x=\Delta z=0$.
	The superscript $+$ denotes the normalization with viscous units.
	The profiles shown in figure~\ref{fig2a} share similar variation tendencies: starting from zero, approaching a negative peak ($-T_{\rm n}$), raising to a positive peak ($T_{\rm p}$), and finally converging to zero again.
	The zero values of $T_{u'\to\tau_\mathrm{w}}$ observed at large time lags are physically reasonable, since the causal signals can only influence the $\tau_\mathrm{w}$-events within finite temporal windows.
	As $y^+$ increases, it takes a longer time duration for the causal information carried by $u'$ propagating to the wall, and the magnitudes of the peaks ($T_{\rm n}$ and $T_{\rm p}$) are subsequently reduced, which is consistent with the inclination characteristic of large-scale motions and further supports the applicability of LKIF in revealing the causal mechanism of the TSD generation.
	Figure~\ref{fig2b} displays the variations of $T_{\rm n}$ and $T_{\rm p}$ and their corresponding time of arrival ($-\Delta t_{\rm n}^+$ and $-\Delta t_{\rm p}^+$) across the whole wall layer.
	Similar to the examples shown in figure~\ref{fig2a}, $T_{\rm p}$ is larger than $T_{\rm n}$ at all wall-normal positions, suggesting the superiority of positive causality exerted on the TSD generation.
	As $y^+$ increases, both $T_{\rm p}$ and $T_{\rm n}$ decay rapidly and become negligible at the edge of the wall layer.
	Meanwhile, $\Delta t_{\rm p}^+$ and $\Delta t_{\rm n}^+$ increase slowly in the inner layer ($y^+<10$) and reach $\Delta t ^+\approx50$ at the edge, with $\Delta t_{\rm p}^+<\Delta t_{\rm n}^+$ regardless of the wall-normal positions.

	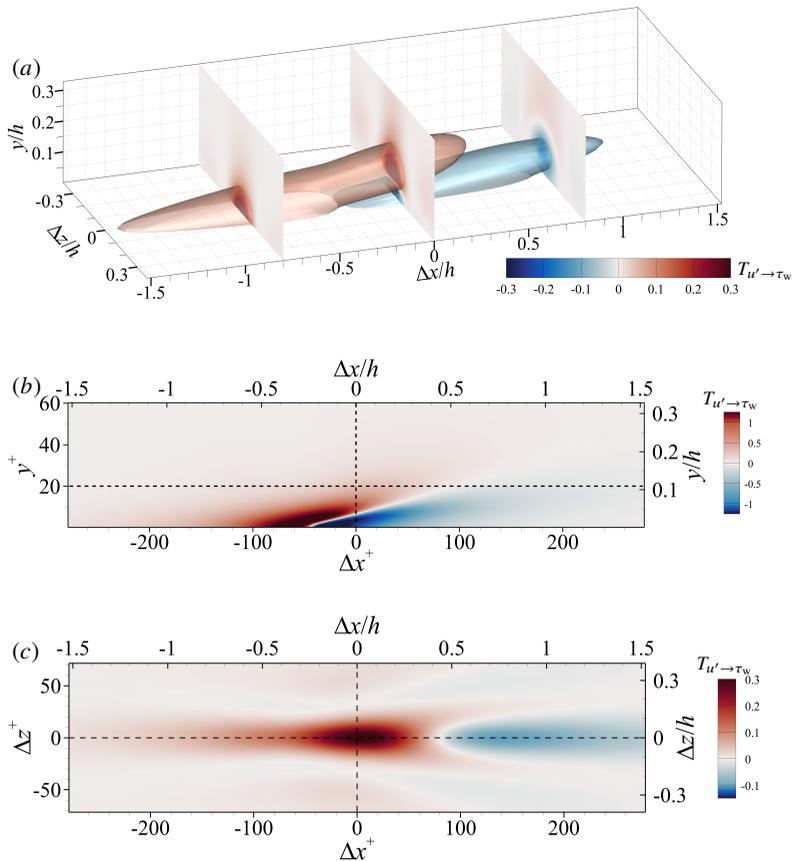
\begin{figure}
		\centering
		\subfigure{
			\begin{overpic}[width=4in]{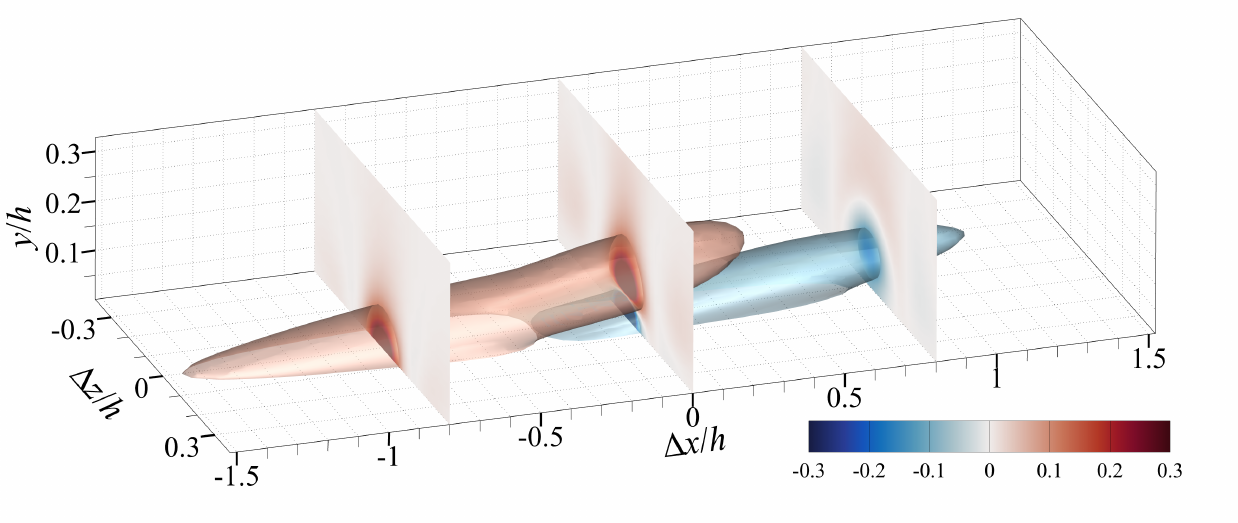}
				\put(1,32){($a$)}
				\put(95.5,6){\scalebox{0.8}{$T_{u'\to\tau_\mathrm{w}}$}}
			\end{overpic}\label{fig3a}}
		\subfigure{
			\begin{overpic}[width=4in]{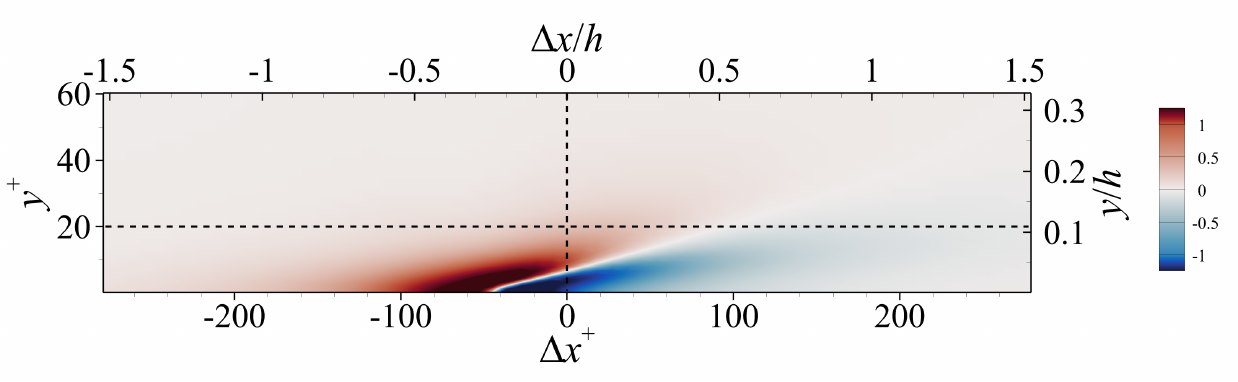}
				\put(1,24.5){($b$)}
				\put(91,23.1){\scalebox{0.8}{$T_{u'\to\tau_\mathrm{w}}$}}
			\end{overpic}\label{fig3b}}
		\subfigure{
			\begin{overpic}[width=4in]{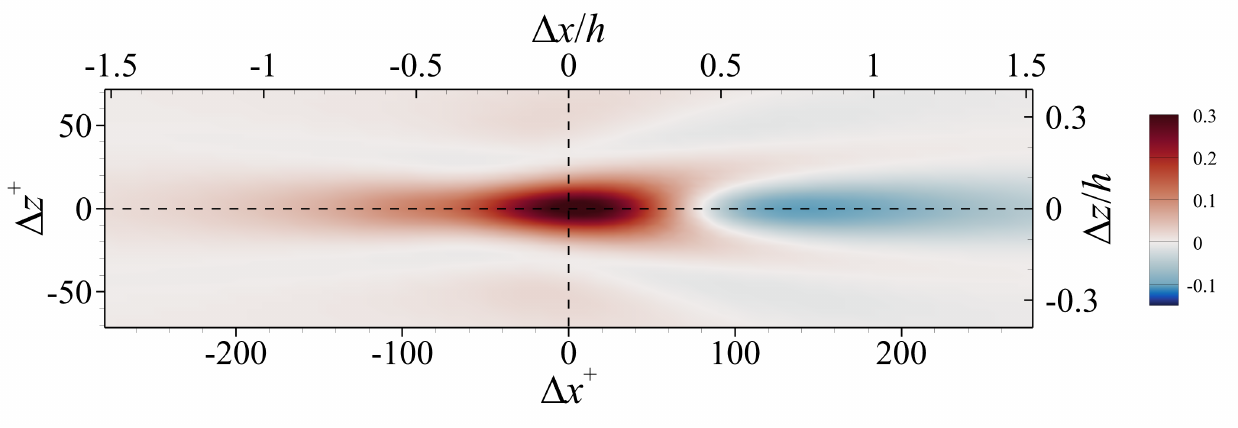}
				\put(1,28){($c$)}
				\put(90.2,26.5){\scalebox{0.8}{$T_{u'\to\tau_\mathrm{w}}$}}
			\end{overpic}\label{fig3c}}
		\caption{Spatial distribution of $T_{u'\to\tau_\mathrm{w}}$ at the time lag $\Delta t^+=-5$.
			($a$) The three-dimensional view of $T_{u'\to\tau_\mathrm{w}}$.
			The positive and negative structures are displayed by the isosurfaces of $T_{u'\to\tau_\mathrm{w}}=\pm0.1$.
			Three slices show contours of $T_{u'\to\tau_\mathrm{w}}$ at $\Delta x=-0.8h$, $0h$, and $0.8h$, respectively.
			($b$) Colored contour of $T_{u'\to\tau_\mathrm{w}}$ at the plane $\Delta z=0$.
			($c$) Colored contour of $T_{u'\to\tau_\mathrm{w}}$ at the plane $y^+=20$.}
		\label{fig3}
	\end{figure}
	
	Analogous to the definition of coherent structures given by \cite{Robinson1991}, here we define the causal structure as `\textit{a statistical consecutive region of the flow over which at least one fundamental flow variable exhibits causality with another variable}'.
	Since $T_{u'\to\tau_\mathrm{w}}\left(\Delta x,y,\Delta z,\Delta t\right)$ is a function of space and time, the temporal evolutions of the causal structures are animated and provided in supplementary movie 1.	
	The movie illustrates that the causal structures propagate from upstream to downstream, and the spatial distributions of the causal structures exhibit no essential difference at different time lags.
	Hereafter, we mainly focus on their spatial characteristics by taking $\Delta t^+=-5$ as an example, when $T_{u'\to\tau_\mathrm{w}}$ at $y^+=20$ reaches its positive peak.
	
	Figure~\ref{fig3a} presents a snapshot of the three-dimensional $u'$-causal structures at $\Delta t^+=-5$ by using isosurfaces of $T_{u'\to\tau_\mathrm{w}}=\pm0.1$ for better visualizations.
	Three $y$-$z$ cross-sections of $T_{u'\to\tau_\mathrm{w}}$ are also added.
	It can be seen that the $u'$-causal structures are streamwise elongated and spanwise symmetric, which geometrically resembles the streamwise streaks~\citep{Kim1987}.
	Differently, the $u'$-causal structures consist of a positive and a negative part, with the former lying upstream and appearing later in time.
	The connected positive part is characterised by $\lambda_x^+\approx360$ and $\lambda_z^+\approx55$ which denote the length and width, respectively, of its bounding box.
	The scales of the negative part are in the same orders but slightly smaller than the positive one.
	Recalling the classical streamwise streaks whose characteristic scales are typically $\lambda_x^+\approx1000$ and $\lambda_z^+\approx100$~\citep{Kline1967,Kim1987,Hwang2013,Giovanetti2017}, the $u'$-causal structures (combining the positive and negative parts) identified by LKIF exhibit more compact spatial distributions, being one half shorter and narrower than those of streamwise streaks.
	To illustrate the $u'$-causal structure in the wall-normal direction, figure~\ref{fig3b} plots the contours of $T_{u'\to\tau_\mathrm{w}}$ in the $x$-$y$ plane with $\Delta z=0$.
	Both the positive and negative parts of the structure are seen to be attached to the wall, with an inclination angle of $\sim8^\circ$ to the streamwise direction.
	The wall-attached feature is closely consistent with Townsend's attached eddy hypothesis (AEH)~\citep{Townsend1976}, and the inclination angle is the same as that of near-wall streamwise streaks~\citep{Jeong1997}.
	Figure~\ref{fig3c} shows the contours of $T_{u'\to\tau_\mathrm{w}}$ in the $x$-$z$ plane at $y^+=20$.
	The positive part bifurcates and envelops the connecting edges of the negative one, and both are symmetric in the spanwise direction.
	
	Based on the interpretation of LKIF~\citep{Liang2014}, the positive and negative parts have distinct influences on the target $\tau_\mathrm{w}$-events.
	The former serves to promote the growth of information entropy contained in the $\tau_\mathrm{w}$-events, exciting stronger fluctuations of $\tau_\mathrm{w}$ and generating more extreme $\tau_\mathrm{w}$-events.	
	By contrast, the negative part has actions to reduce the information entropy, and consequently suppress the formation of extreme $\tau_\mathrm{w}$-events.
	From the perspective of mitigating adverse effects associated with the extreme $\tau_\mathrm{w}$-events (such as noise radiation and structural vibrations), these observations suggest a novel strategy that precisely controls the positive structure within well-defined spatial domains and temporal windows.
	
	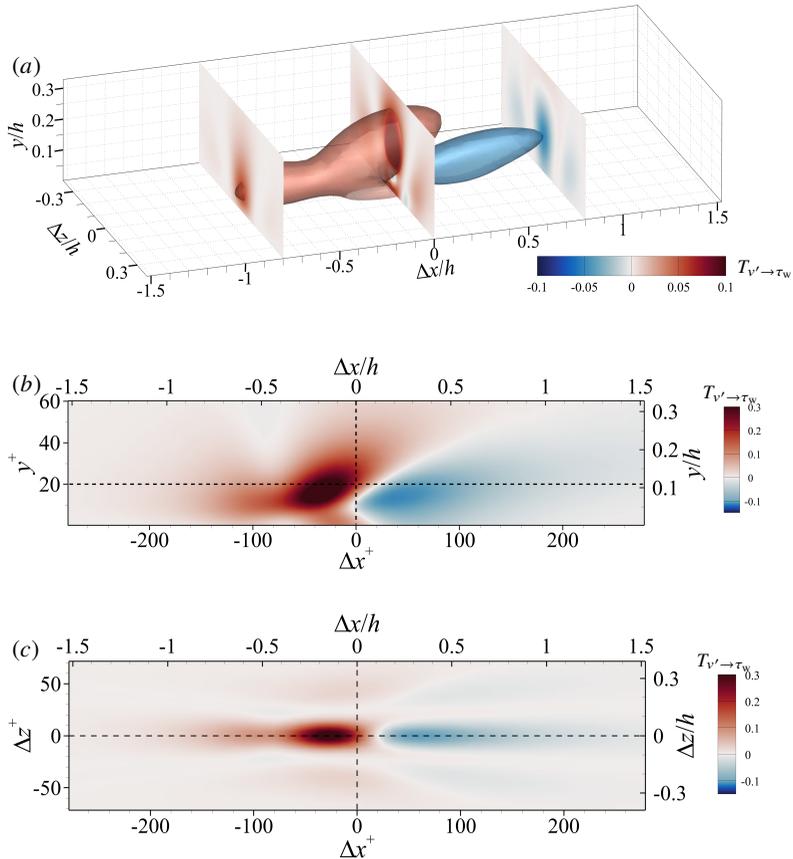
\begin{figure}
		\centering
		\subfigure{
			\begin{overpic}[width=4in]{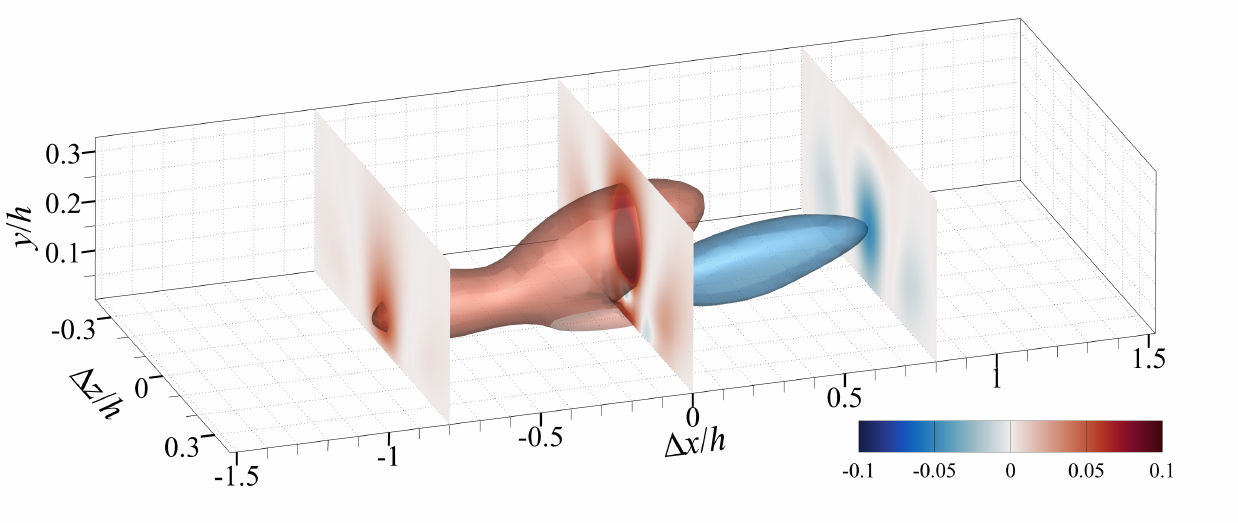}
				\put(1,32){($a$)}
				\put(95.5,6){\scalebox{0.8}{$T_{v'\to\tau_\mathrm{w}}$}}
			\end{overpic}\label{fig4a}}
		\subfigure{
			\begin{overpic}[width=4in]{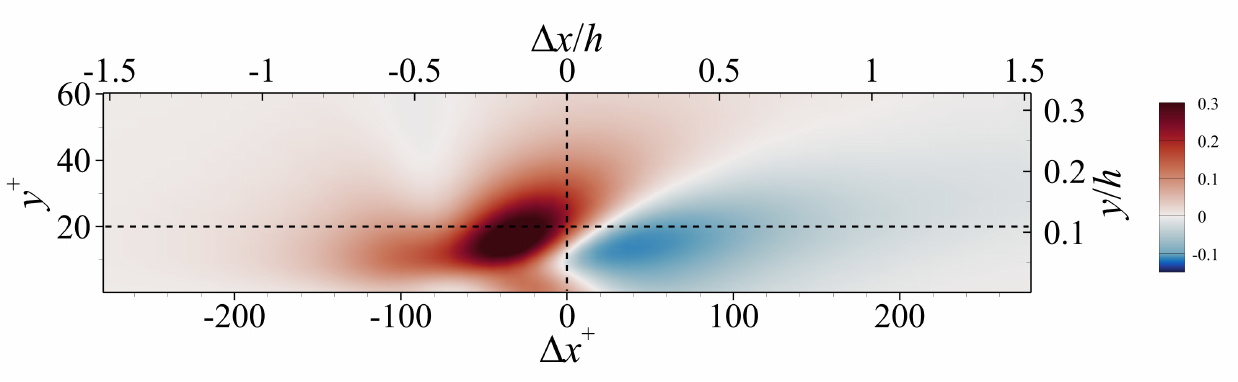}
				\put(1,24.5){($b$)}
				\put(91,23.6){\scalebox{0.8}{$T_{v'\to\tau_\mathrm{w}}$}}
			\end{overpic}\label{fig4b}}
		\subfigure{
			\begin{overpic}[width=4in]{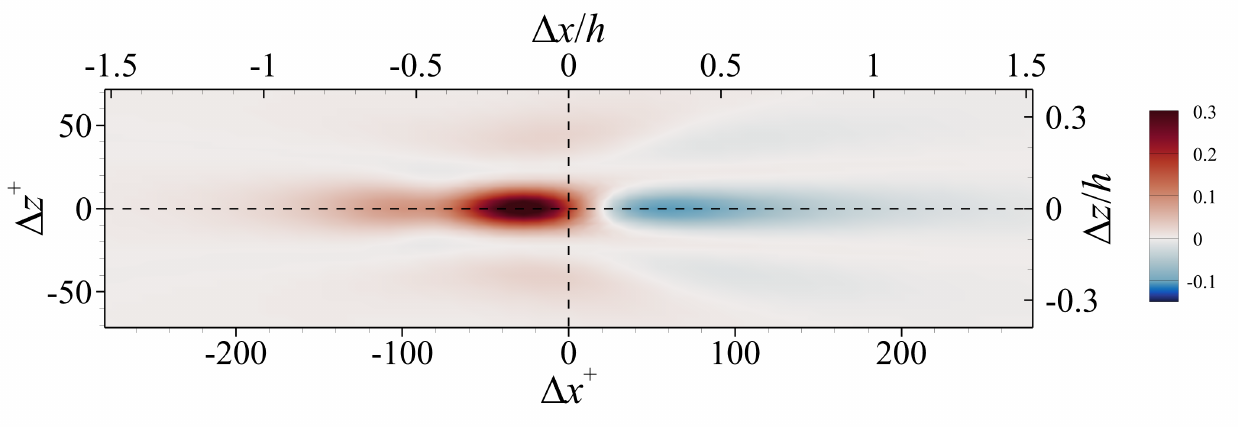}
				\put(1,28){($c$)}
				\put(90.2,26.7){\scalebox{0.8}{$T_{v'\to\tau_\mathrm{w}}$}}
			\end{overpic}\label{fig4c}}
		\caption{Spatial distribution of $T_{v'\to\tau_\mathrm{w}}$ at the time lag $\Delta t^+=-5$.
			($a$) The three-dimensional view of $T_{v'\to\tau_\mathrm{w}}$.
			The positive and negative structures are displayed by the isosurfaces of $T_{v'\to\tau_\mathrm{w}}=\pm0.05$.
			Three slices show contours of $T_{v'\to\tau_\mathrm{w}}$ at $\Delta x=-0.8h$, $0h$, and $0.8h$, respectively.
			($b$) Colored contour of $T_{v'\to\tau_\mathrm{w}}$ at the plane $\Delta z=0$.
			($c$) Colored contour of $T_{v'\to\tau_\mathrm{w}}$ at the plane $y^+=20$.}
		\label{fig4}
	\end{figure}
	
	In a similar way, the spatial distributions of $T_{v'\to\tau_\mathrm{w}}$ are shown in figure~\ref{fig4} to analyze the causal relationships between $v'$ and $\tau_\mathrm{w}$.
	On the whole, the $v'$-causal structures share similarities with the $u'$-causal structures, but exhibiting two distinguishing characteristics.
	Firstly, the $v'$-causal structures have more compact length scales in the wall-parallel directions (see figure~\ref{fig4c}), and the cores of both positive and negative parts are located at higher wall-normal positions (see figure~\ref{fig4b}), exhibiting wall-detached distributions with an inclination angle of $\sim18^\circ$.
	The wall-detached feature is also consistent with the hypothesis of Townsend's attached-eddy model, which is attributed to the impenetrability of the wall preventing $v'$ motions from approaching the wall-surface.
	Secondly, figure~\ref{fig4c} reveals a distinct streamwise positional delay compared with figure~\ref{fig3c}, suggesting the desynchrony of $v'$ and $u'$ motions in exerting influences on the TSD generation.
	The temporal evolutions of $v'$-causal structures are provided in supplementary movie 2.	
	
	\begin{figure}
		\centering
		\subfigure{
			\begin{overpic}[width=4in]{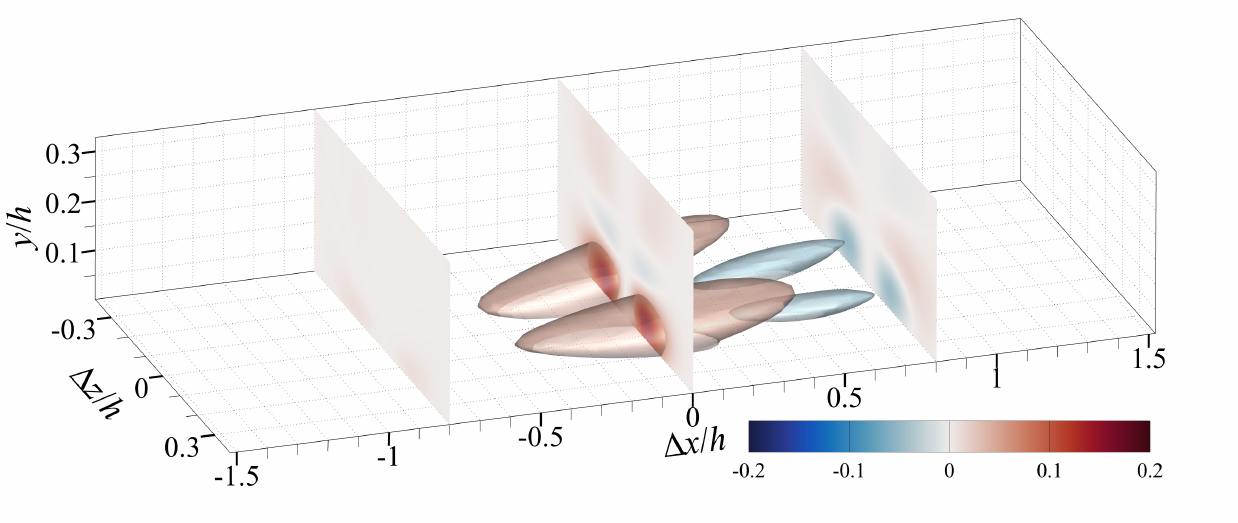}
				\put(1,32){($a$)}
				\put(94.5,6){\scalebox{0.8}{$T_{w'\to\tau_\mathrm{w}}$}}
			\end{overpic}\label{fig5a}}
		\subfigure{
			\begin{overpic}[width=4in]{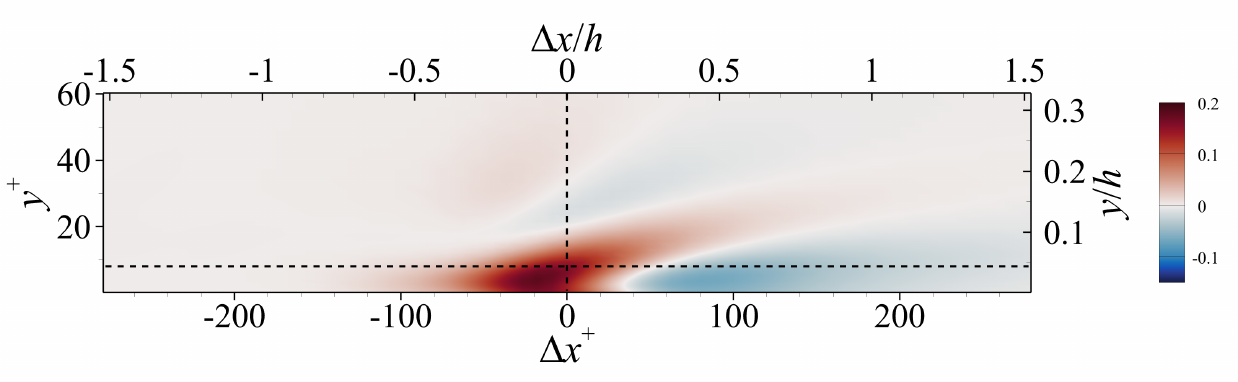}
				\put(1,24.5){($b$)}
				\put(90.6,23.8){\scalebox{0.8}{$T_{w'\to\tau_\mathrm{w}}$}}
			\end{overpic}\label{fig5b}}
		\subfigure{
			\begin{overpic}[width=4in]{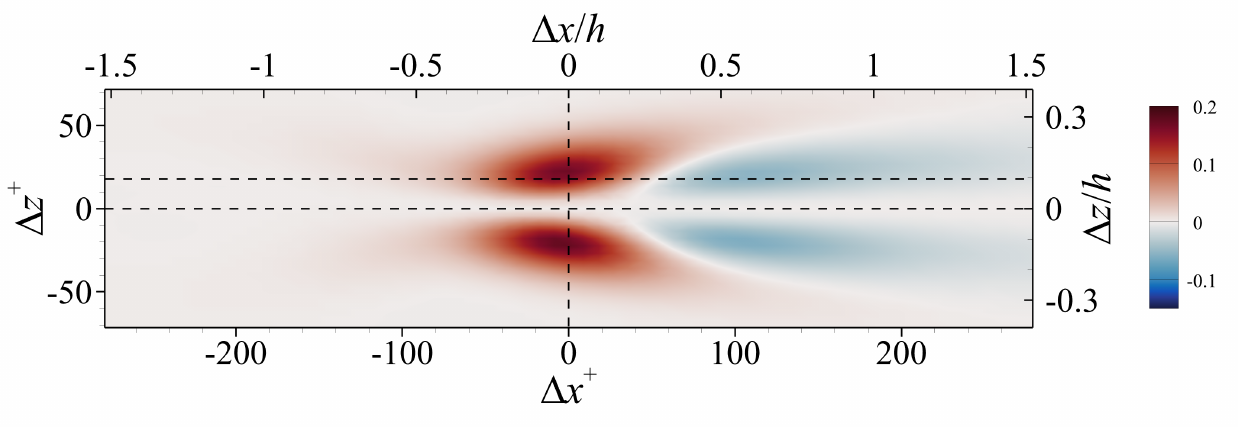}
				\put(1,28){($c$)}
				\put(89.9,27.2){\scalebox{0.8}{$T_{w'\to\tau_\mathrm{w}}$}}
			\end{overpic}\label{fig5c}}
		\caption{Spatial distribution of $T_{w'\to\tau_\mathrm{w}}$ at the time lag $\Delta t^+=-5$.
			($a$) The three-dimensional view of $T_{w'\to\tau_\mathrm{w}}$.
			The positive and negative structures are displayed by the isosurfaces of $T_{w'\to\tau_\mathrm{w}}=\pm0.05$.
			Three slices show contours of $T_{w'\to\tau_\mathrm{w}}$ at $\Delta x=-0.8h$, $0h$, and $0.8h$, respectively.
			($b$) Colored contour of $T_{w'\to\tau_\mathrm{w}}$ at the plane $\Delta z^+=18$.
			The horizontal dashed line denotes $y^+=8$.
			($c$) Colored contour of $T_{w'\to\tau_\mathrm{w}}$ at the plane $y^+=8$.
			The upper horizontal dashed line denotes $\Delta z^+=18$.}
		\label{fig5}
	\end{figure}
	
	Figure~\ref{fig5} shows the $w'$-causal structure identified by $T_{w'\to\tau_\mathrm{w}}$.
	A prominent characteristic is that the positive and negative parts of $w'$-causal structures are respectively bipartite and symmetric with respect to the zero-causality plane ($\Delta z=0$).
	Compared with $u'$-causal structures, high-causality zones of $w'$-causal structures are closer to the wall and have smaller magnitudes.
	Based on this observation, representative cross-sections are selected at $\Delta z^+=18$ ($x$-$y$ slice, figure~\ref{fig5b}) and $y^+=8$ ($x$-$z$ slice, figure~\ref{fig5c}) for better visualization.
	As displayed in figure~\ref{fig5b}, the $w'$-causal structures are attached to the wall with the same inclination angle ($\sim8^\circ$) as the $u'$-causal structures.
	Figure~\ref{fig5c} exhibits the similar bifurcation of positive parts enveloping the upstream edge of negative parts.
	The temporal evolutions of $w'$-causal structures are provided in supplementary movie 3.
	
	To summarize, the causal structures identified by LKIF explicitly reveal the well-organized spatial-temporal characteristics of flow motions responsible for the TSD generation, such as wall-attachment behaviors, length scales, inclination angles, and the key positive-negative regularities.
	These characteristics are notably important for understanding TSD generation and designing drag reduction strategies.

	\subsection{Comparison with classical correlation analysis}\label{sec4.2}

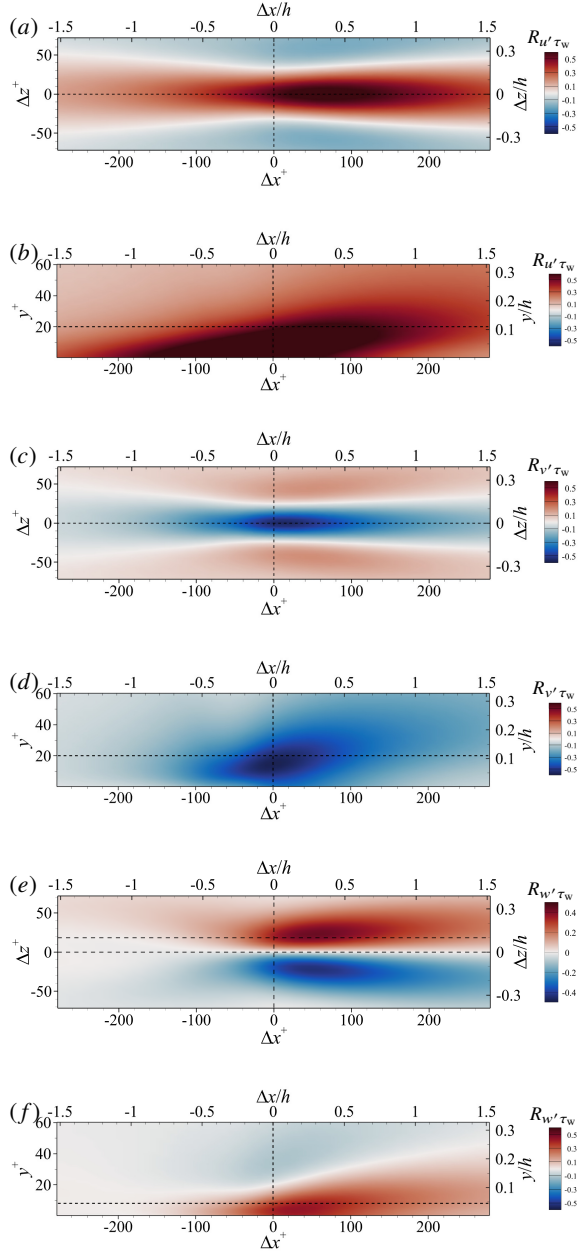
\begin{figure}
		\centering
		\subfigure{
			\begin{overpic}[width=3in]{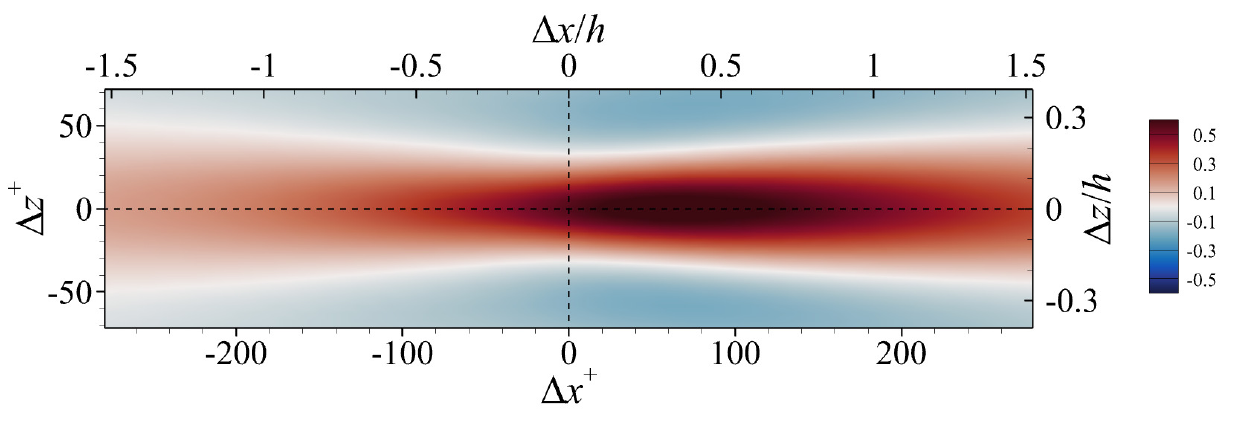}
				\put(0,28){($a$)}
				\put(89.8,26.5){\scalebox{0.8}{$R_{u'\tau_\mathrm{w}}$}}
			\end{overpic}\label{fig6a}}
		\subfigure{
			\begin{overpic}[width=3in]{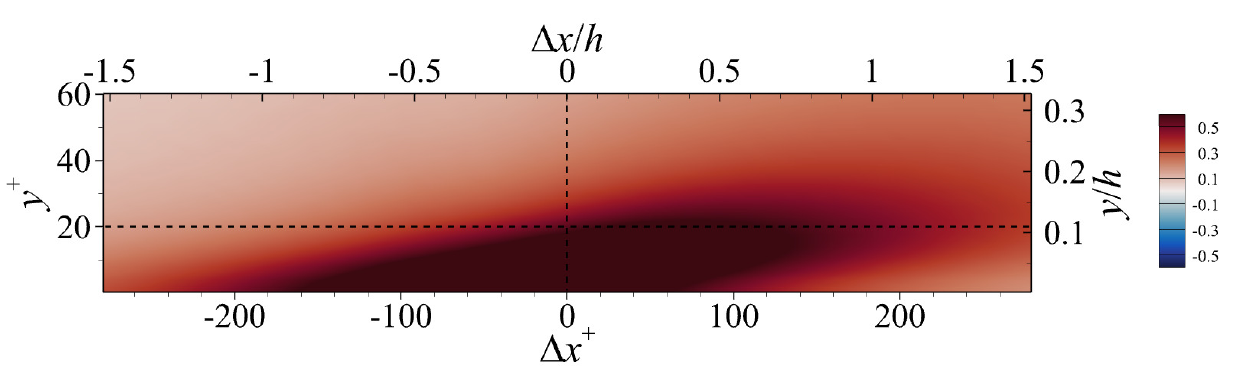}
				\put(0,24){($b$)}
				\put(90.7,23.1){\scalebox{0.8}{$R_{u'\tau_\mathrm{w}}$}}
			\end{overpic}\label{fig6b}}
		\subfigure{
			\begin{overpic}[width=3in]{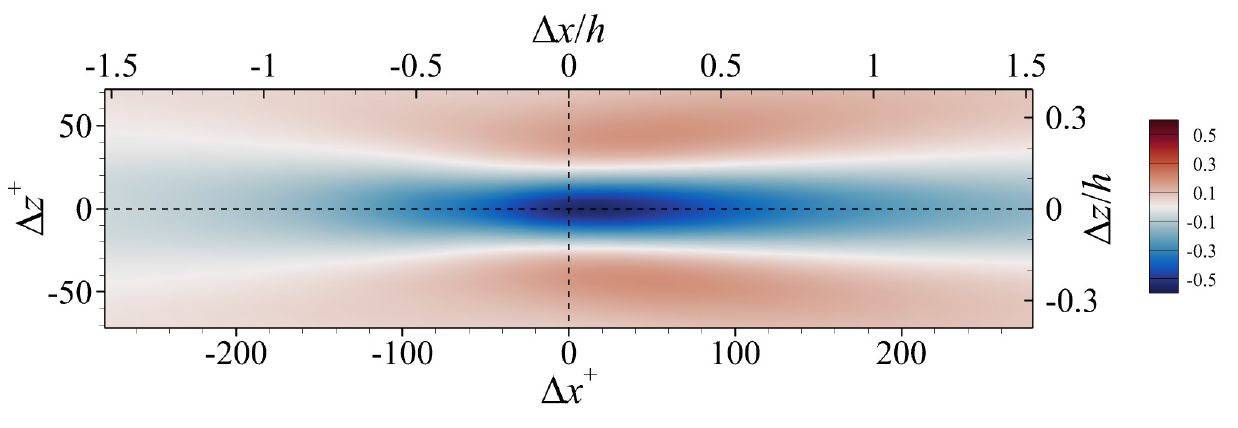}
				\put(0,28){($c$)}
				\put(89.8,26.6){\scalebox{0.8}{$R_{v'\tau_\mathrm{w}}$}}
			\end{overpic}\label{fig6c}}
		\subfigure{
			\begin{overpic}[width=3in]{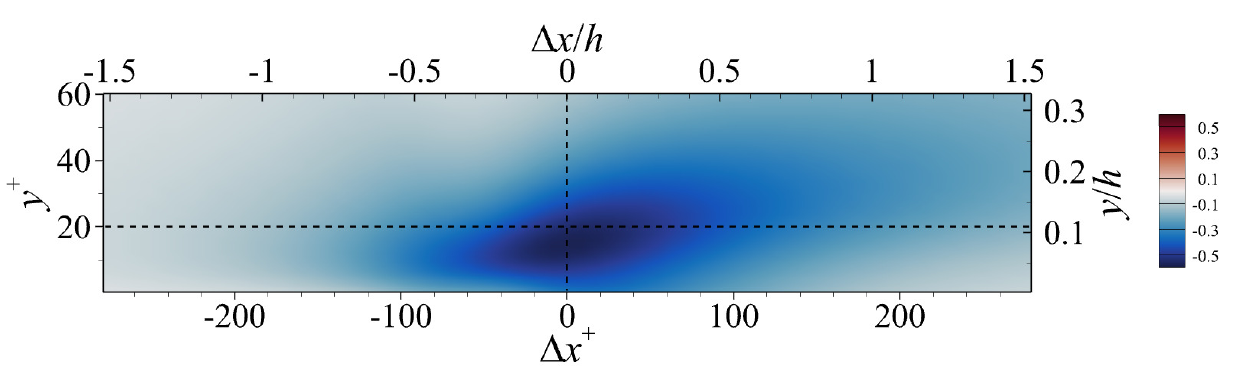}
				\put(0,24){($d$)}
				\put(90.7,23.2){\scalebox{0.8}{$R_{v'\tau_\mathrm{w}}$}}
			\end{overpic}\label{fig6d}}
		\subfigure{
			\begin{overpic}[width=3in]{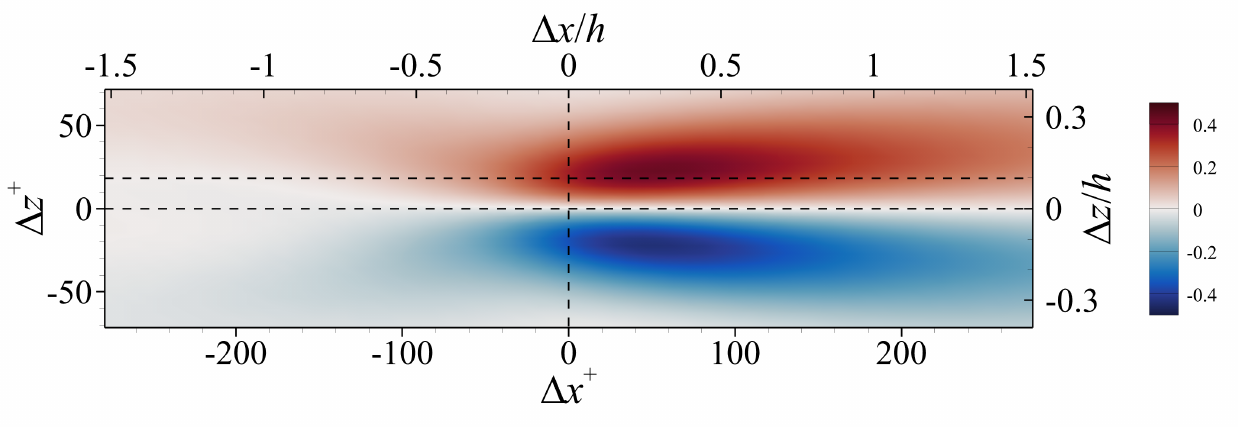}
				\put(0,28){($e$)}
				\put(89.8,27.6){\scalebox{0.8}{$R_{w'\tau_\mathrm{w}}$}}
			\end{overpic}\label{fig6e}}
		\subfigure{
			\begin{overpic}[width=3in]{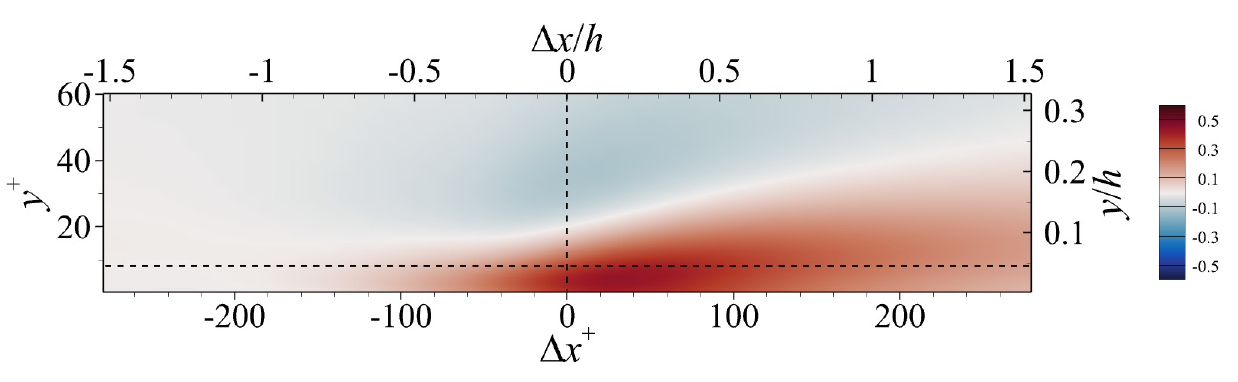}
				\put(0,24){($f$)}
				\put(90.2,23.8){\scalebox{0.8}{$R_{w'\tau_\mathrm{w}}$}}
			\end{overpic}\label{fig6f}}
		\caption{Spatial distributions of correlation between velocity components and the target $\tau_\mathrm{w}$ at the time lag $\Delta t^+=-5$.
		($a$) $R_{u'\tau_\mathrm{w}}$ at $y^+=20$, ($b$) $R_{u'\tau_\mathrm{w}}$ at $\Delta z=0$, ($c$) $R_{v'\tau_\mathrm{w}}$ at $y^+=20$, ($d$) $R_{v'\tau_\mathrm{w}}$ at $\Delta z=0$, ($e$) $R_{w'\tau_\mathrm{w}}$ at $y^+=8$, and ($f$) $R_{w'\tau_\mathrm{w}}$ at $\Delta z^+=18$.}
		\label{fig6}
	\end{figure}

	As we know, correlation does not necessarily imply causality, but causality definitely includes correlation.
	To provide a straight comparison between the correlation and causal analysis, we calculate the correlation fields ($R_{u'\tau_\mathrm{w}}$, $R_{v'\tau_\mathrm{w}}$ and $R_{w'\tau_\mathrm{w}}$) between velocity components ($u'$, $v'$ and $w'$) and the target $\tau_\mathrm{w}$, and plot their cross-sectional distributions in figure~\ref{fig6} using the same calculation process and selected time lag as the LKIF fields shown in figures~\ref{fig3}$\sim$\ref{fig5}.
	$R_{\phi\tau_\mathrm{w}}$ is computed by $R_{\phi\tau_\mathrm{w}}=C_{\phi\tau_\mathrm{w}}/\left(\phi_\mathrm{rms}\tau_\mathrm{w,rms}\right)$, where $C_{\phi\tau_\mathrm{w}}$ is the covariance between $\phi$ and $\tau_\mathrm{w}$, and the subscript `rms' indicates the root mean square.
	
	In the vicinity of the target $\tau_\mathrm{w}$, as seen in figures~\ref{fig6a}$\sim$\ref{fig6d}, the $u'$ and $v'$ correlations are positively and negatively correlated with the TSD generation, respectively.
	Both of them exhibit streamwise-elongated and spanwise-symmetric features.
	From figures~\ref{fig6e} and~\ref{fig6f}, we find that the $w'$ correlations are also streamwise-elongated but anti-symmetric in the spanwise direction.
	Meanwhile, all components of the correlations are inclined in the wall-normal direction, whereas $u'$ and $w'$ attach to the wall, and $v'$ detaches from the wall.
	
	Although the structures identified by the correlation and causal analysis share qualitative similarities, they differ in two key aspects.
	Firstly, the causal structures only lie in the regions where correlations are significant.
	There are substantial regions where correlations are present but without causality, and consequently the scales of correlation structures are notably larger than those of causal structures.
	These differences immediately support that `correlation does not imply causation'.
	Secondly, the sign of LKIF is able to separate the causal structures that generate more extreme $\tau_\mathrm{w}$-events or less (as discussed in the preceding section), while the sign of correlation just reflects the synchrony between the velocity components and the targeted $\tau_\mathrm{w}$, which fails to separate the influences on the extreme $\tau_\mathrm{w}$-events.
	In this respect, the causal structures demonstrate their exceptional advantages in physical interpretability of the TSD generation.
	
\begin{figure}
		\centering
		\includegraphics[width=2.5in]{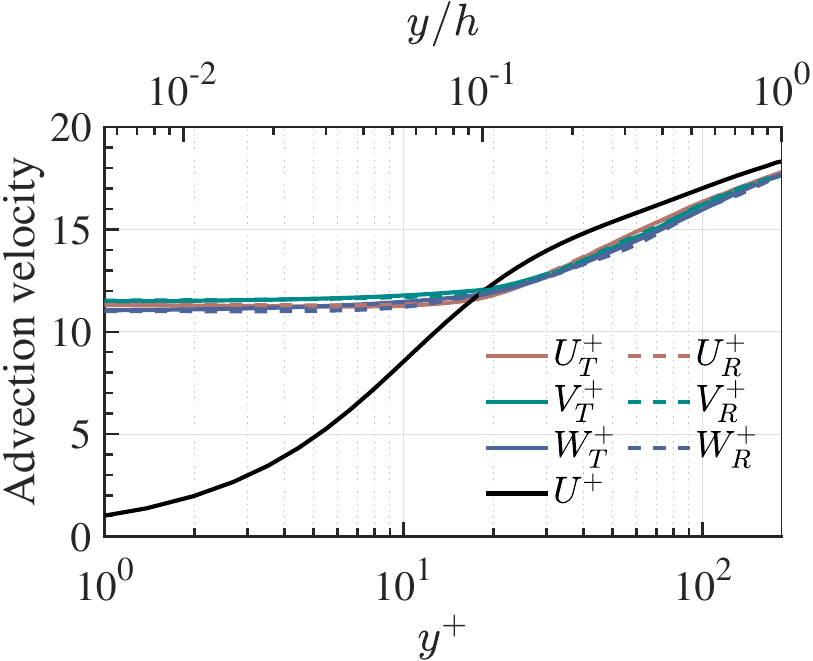}
		\caption{Profiles of advection velocity of causal structures ($U_T^+$, $V_T^+$, and $W_T^+$) and correlations ($U_R^+$, $V_R^+$, and $W_R^+$), as well as the mean streamwise velocity ($U^+$).}
		\label{fig7}
	\end{figure}

	Analogous to causal structures, apart from space, correlation is also a function of time, relying on the time lag ($\Delta t$) of $u'$ series relative to $\tau_\mathrm{w}$ series.
	As the time lag increases, both the causal and correlation structures advect downstream, influencing the TSD generation within limited temporal windows (see the supplementary movie 4 and movie 5).
	Their advection velocities are of fundamental interest, and we therefore calculate the velocities at each wall-normal plane by regarding the velocity of peak point as the indicator of advection velocity, i.e. $U_\mathrm{adv}(y)=\delta x_\mathrm{p}(y)/\delta\Delta t$.
	Here, $\delta x_\mathrm{p}$ is the streamwise displacement of the peak point (positive or negative) of causal or correlation structures within a given time length $\delta\Delta t$.
	Advection velocities are plotted in figure~\ref{fig7} together with the speed of the mean flow ($U^+$).
	Advection velocities of the causal structures ($U_T^+$, $V_T^+$, and $W_T^+$) represent the speeds of information propagation, and the advection velocities of the correlation structures ($U_R^+$, $V_R^+$, and $W_R^+$) represent the speeds of perturbations related to the wall shear stress~\citep{Kim1993}.
	As shown in figure~\ref{fig7}, the differences among the advection velocities ($U_T^+$, $V_T^+$, $W_T^+$, $U_R^+$, $V_R^+$, and $W_R^+$) are insignificant, as the information and perturbations carried by different velocity components are maintained in the same physical events to generate the variations of TSD.
	It is not a coincidence that the advection velocities of correlations collapse well with the causal structures, since LKIF of each velocity component is mathematically estimated by algebraic operations of the auto- and cross-correlations.
	In a physical view, it suggests that the causal and correlation structures are propagating at the same speeds.
	In the viscous sublayer ($y^+<5$), both the causal and correlation structures advect at virtually constant velocities and much faster than the mean flow, implying that the propagation of information and perturbations does not necessarily rely on the mean flow.
	The structures advect faster until $y^+\approx16$, from where the situation reverses: the velocity of mean flow exceeds that of structures and keeps this gap until the edge of the wall layer.
	It indicates that the information always possesses a certain wall-ward propagation velocity component so that its advection velocity is lower than the mean flow.
	
	\subsection{Relations between the causal structures and streaks/rolls}\label{sec4.3}
	
	What follows is the question: what flow motions are fundamentally important to form the spatial-temporal distributions of the causal structures.
	It motivates us to analyze the relations of the causal structures with the streamwise streaks and rolls.
	To capture the streamwise streaks and rolls statistically linked to the TSD generation, we compute the conditionally averaged velocity fields based on the extreme $\tau_\mathrm{w}$-events, which are defined as:
	\begin{equation}\label{eq1}
		\langle\bm{u}'\rangle\left(\Delta x,y,\Delta z,\Delta t\right)=\overline{\bm{u}'\left(x+\Delta x,y,z+\Delta z,t+\Delta t\right)}|_{\tau_\mathrm{w}\left(x,z,t\right)>\tau_{\mathrm{w},0.95}},
	\end{equation}
	where $\langle\cdot\rangle$ denotes the conditional average, the overbar means the average over $\left(x,z,t\right)$, $\tau_{\mathrm{w},0.95}$ indicates a probability threshold subject to $\mathbb{P}\left\{\tau_\mathrm{w}>\tau_{\mathrm{w},0.95}\right\}=0.05$,
	and $\mathbb{P}$ is the probability.
	The bold notation $\langle\bm{u}'\rangle$ represents the vector of three conditionally averaged velocity components, i.e.~$\langle u'\rangle$, $\langle v'\rangle$, and $\langle w'\rangle$.
	Proper adjustment (0.02$\sim$0.10) of the threshold value 0.05 does not essentially change the resulting $\langle\bm{u}'\rangle$.
	
	\begin{figure}
		\centering
		\subfigure{
			\begin{overpic}[width=3.5in]{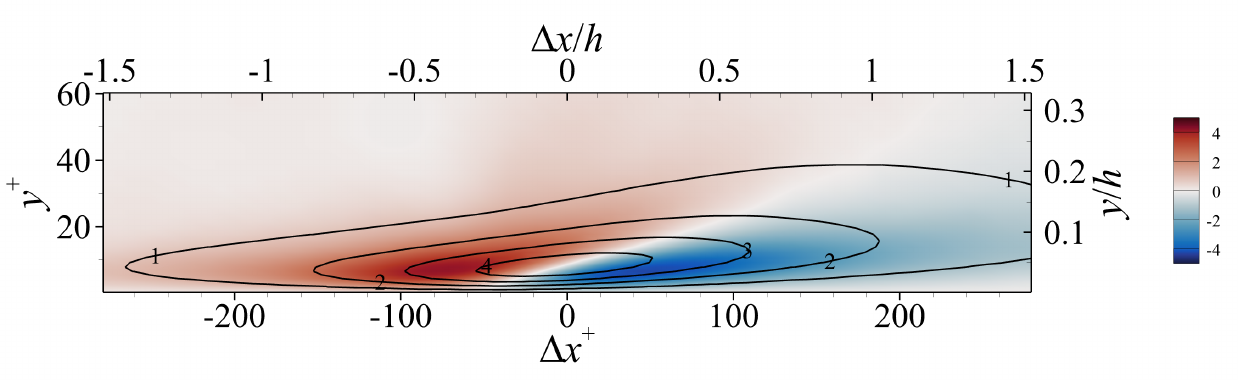}
				\put(1,24.5){($a$)}
				\put(90.6,23.6){\scalebox{0.8}{$\partial\langle u'\rangle/\partial x$}}
			\end{overpic}\label{fig8a}}
		\subfigure{
			\begin{overpic}[width=3.5in]{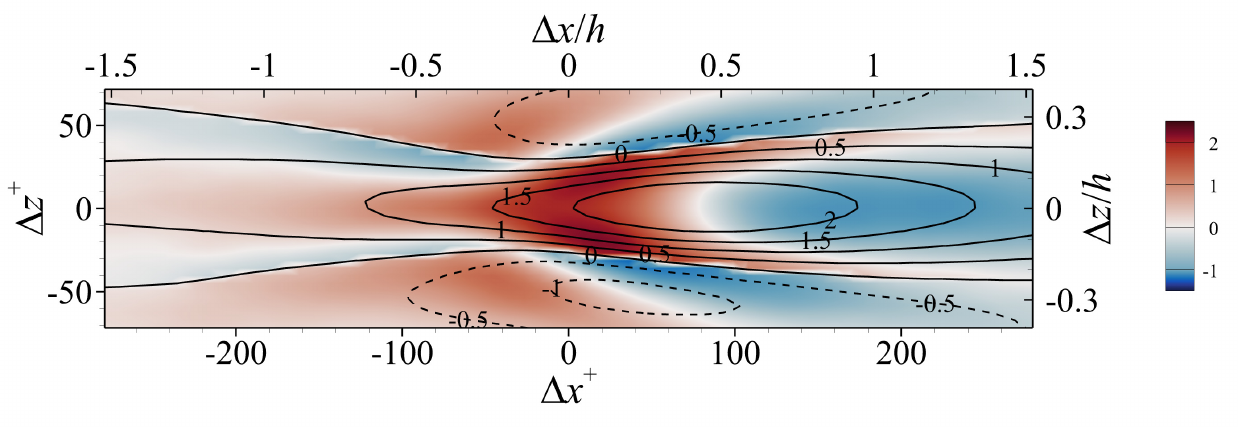}
				\put(1,28){($b$)}
				\put(89,27){\scalebox{0.8}{$\partial|\langle u'\rangle|/\partial x$}}
			\end{overpic}\label{fig8b}}
		\subfigure{
			\begin{overpic}[width=2.6in]{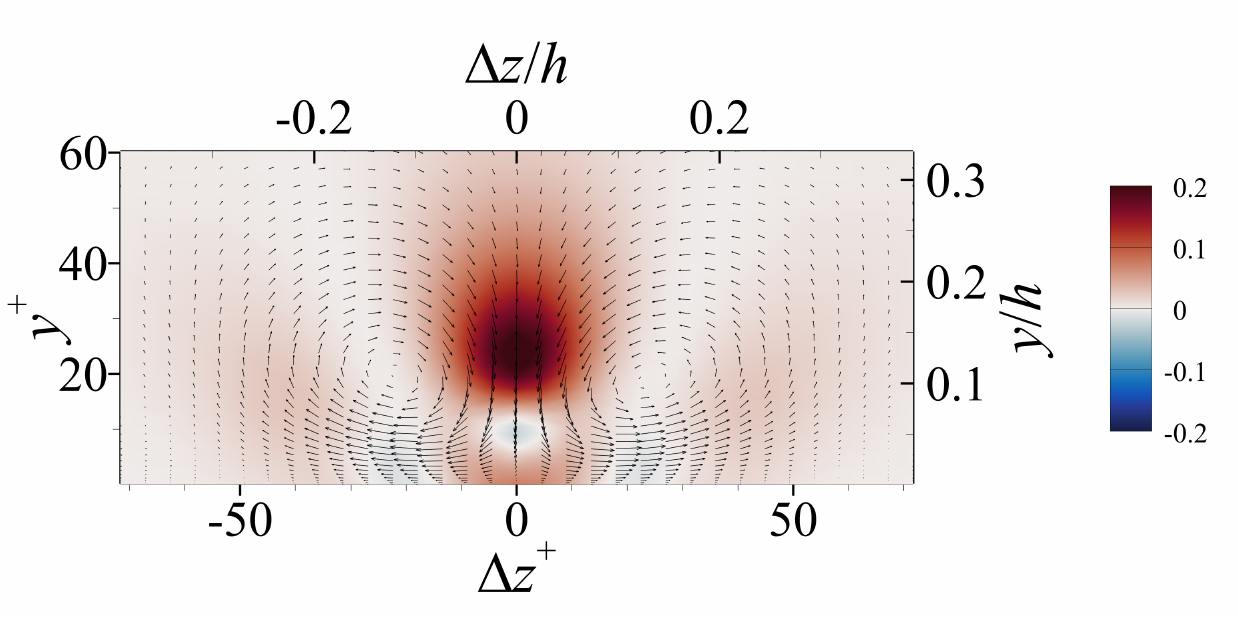}
				\put(1,42){($c$)}
				\put(85.8,37){\scalebox{0.8}{$T_{v'\to\tau_\mathrm{w}}$}}
			\end{overpic}\label{fig8c}}
		\subfigure{
			\begin{overpic}[width=2.6in]{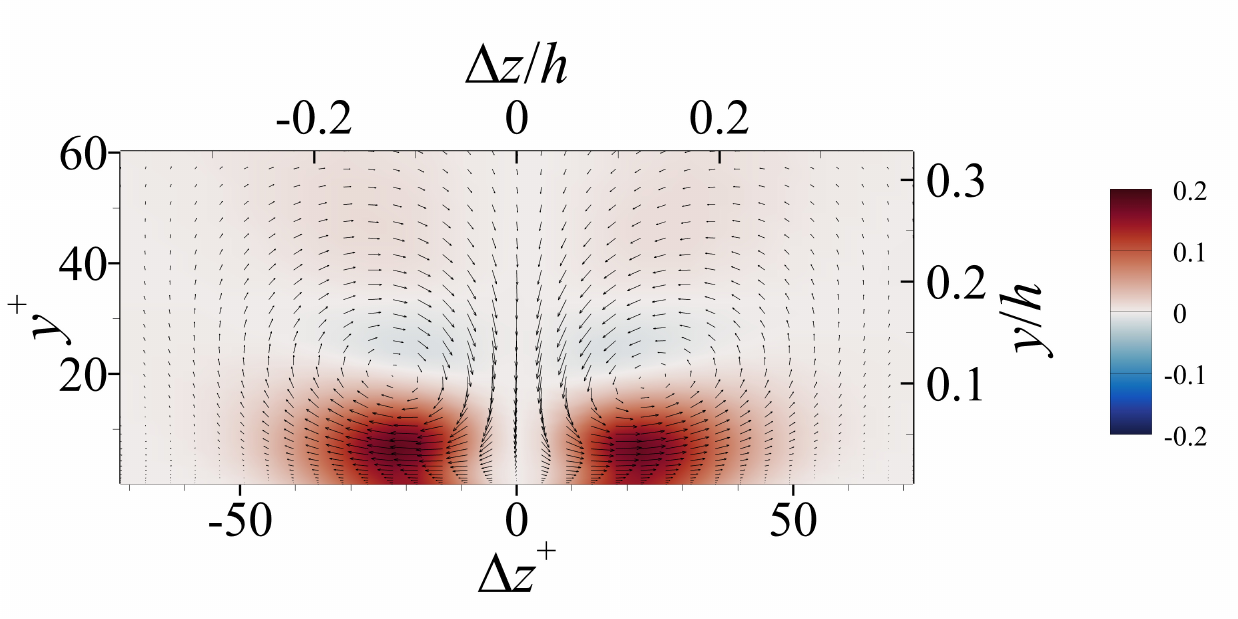}
				\put(1,42){($d$)}
				\put(85,37){\scalebox{0.8}{$T_{w'\to\tau_\mathrm{w}}$}}
			\end{overpic}\label{fig8d}}
		\subfigure{
			\begin{overpic}[width=3.5in]{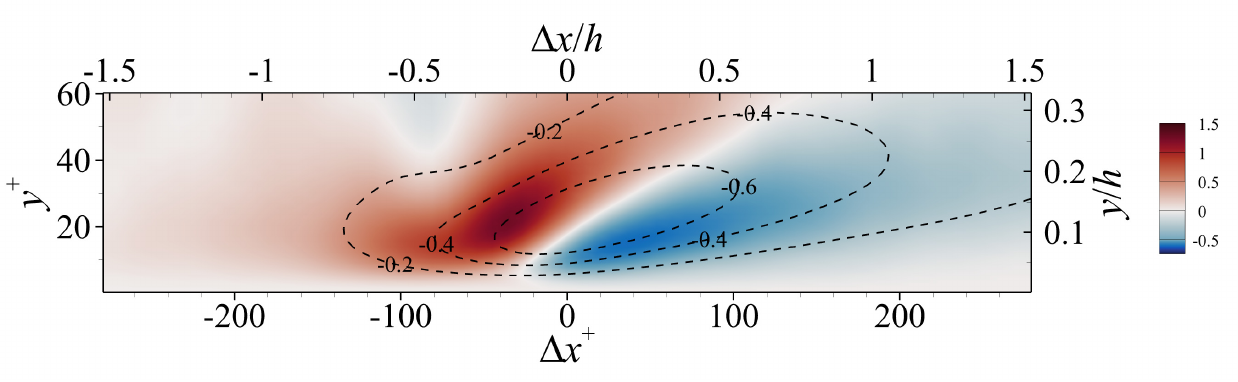}
				\put(1,24.5){($e$)}
				\put(88.5,23){\scalebox{0.8}{$\partial|\langle v'\rangle|/\partial x$}}
			\end{overpic}\label{fig8e}}
		\subfigure{
			\begin{overpic}[width=3.5in]{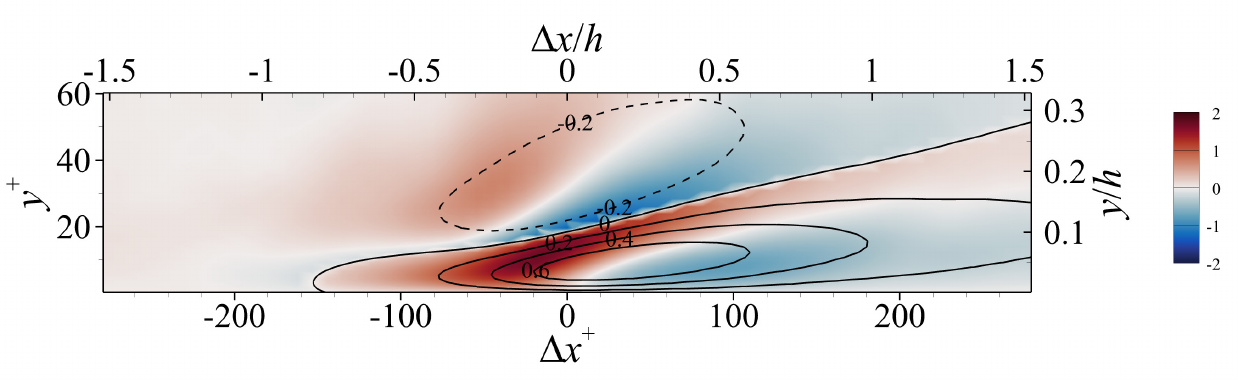}
				\put(1,24.5){($f$)}
				\put(89,23.8){\scalebox{0.8}{$\partial|\langle w'\rangle|/\partial x$}}
			\end{overpic}\label{fig8f}}
		\caption{Distributions of conditionally averaged fields at the time lag $\Delta t^+=-5$: ($a$, $b$) $\langle u'\rangle$ at ($a$) $\Delta z=0$ and ($b$) $y^+=20$ as contour lines, with its values marked on the lines;
			($c$, $d$) $\langle v'\rangle$ coupled with $\langle w'\rangle$ at $\Delta x=0$ as inserted vectors;
			($e$) $\langle v'\rangle$ at $\Delta z=0$ as contour lines, with its values marked on the lines;
			($f$) $\langle w'\rangle$ at $\Delta z^+=18$ as contour lines, with its values marked on the lines.
			The colored contour background of each panel represents: 
			($a$) $\partial\langle u'\rangle/\partial x$ (equivalent to $\partial|\langle u'\rangle|/\partial x$);
			($b$) $\partial|\langle u'\rangle|/\partial x$;
			($c$) $T_{v'\to\tau_\mathrm{w}}$;
			($d$) $T_{w'\to\tau_\mathrm{w}}$;
			($e$) $\partial|\langle v'\rangle|/\partial x$;
			($f$) $\partial|\langle w'\rangle|/\partial x$, respectively.
		}
		\label{fig8}
	\end{figure}
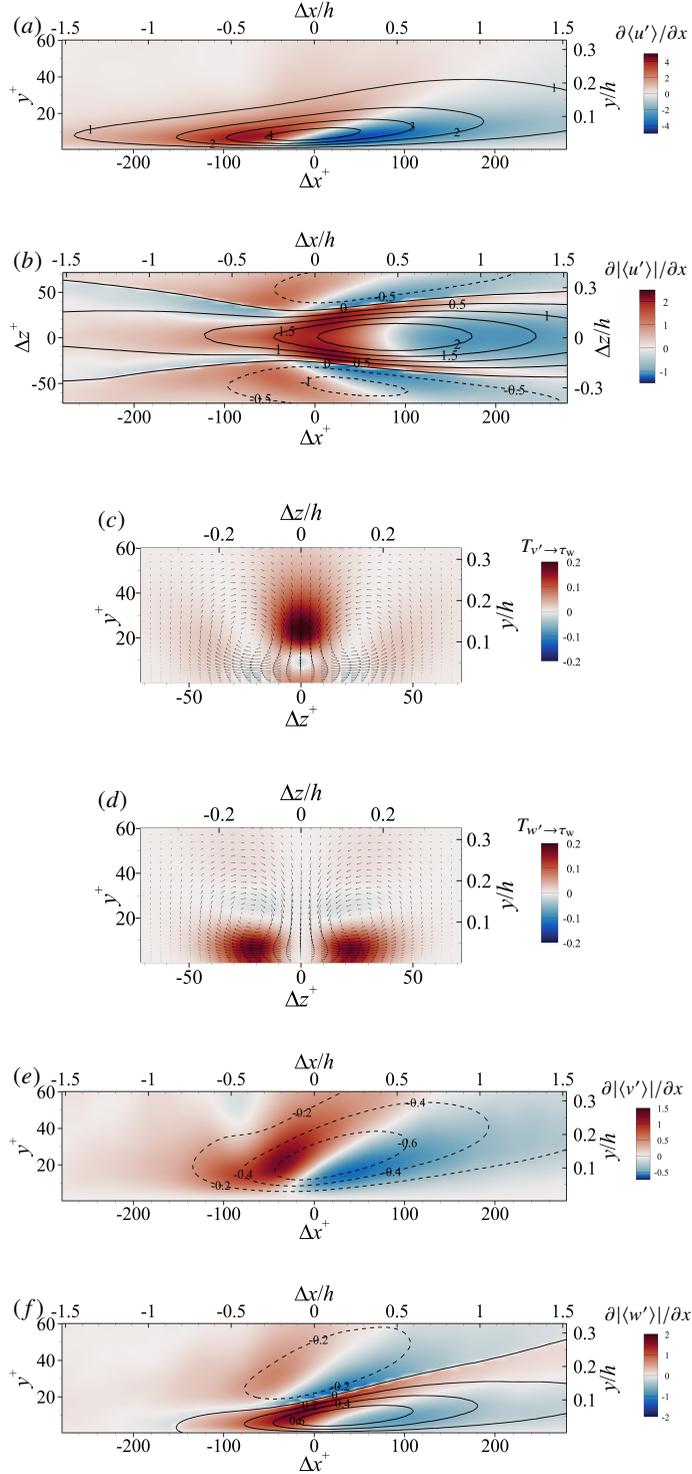
	
	The black contour lines shown in figures~\ref{fig8a} and~\ref{fig8b} explore the distributions of $\langle u'\rangle$ in the $x$-$y$ and $x$-$z$ planes, respectively.
	The streamwise-elongated and normal-inclined feature of high-speed streaks is clearly captured by conditional $\langle u'\rangle$ shown in figure~\ref{fig8a}, and figure~\ref{fig8b} displays that the primary high-speed streak (solid lines) is flanked by low-speed streaks (dashed lines) in the spanwise direction.
	Compared with figures~\ref{fig3b} and~\ref{fig3c}, the conditional streaks appropriately envelop the combined $u'$-causal structures, indicating their unrevealed intrinsic relations, even though the conditional streaks do not exhibit the positive-negative alternation feature along the streamwise direction.
	Considering that the $\langle u'\rangle$ accelerates and decelerates successively along the streamwise direction, we plot the field of $\partial\langle u'\rangle/\partial x$ as the colored contour surfaces in figure~\ref{fig8a}.
	As can be seen, the positive part, the negative part, and their dividing line of $\partial\langle u'\rangle/\partial x$ collapse satisfyingly well with the counterparts of $T_{u'\to\tau_\mathrm{w}}$ shown in figure~\ref{fig3b}.
	As we know, velocity fluctuations imply the offsets away from the mean flow.
	In this sense, the positive $\partial\langle u'\rangle/\partial x$ indicates that the value of $\langle u'\rangle$ is moving away from the mean and approaching its maxima, and vice versa for the negative one.
	This completely consists with the interpretation of LKIF that the positive LKIF promotes the extreme $\tau_\mathrm{w}$-events, while the negative LKIF suppresses that.
	
	Above consistency is also appropriate for the primary high-speed streak shown in figure~\ref{fig8b}.
	However, it seems to be inverse if we turn eyes on the spanwise-flanking low-speed streaks, where $\langle u'\rangle<0$ and $\partial\langle u'\rangle/\partial x$ changes its sign from negative to positive along the streamwise direction, leading to minima of $\langle u'\rangle$.
	We argue that this is not contradictory to the observation that $T_{u'\to\tau_\mathrm{w}}$ still varies from positive to negative at the position of low-speed streaks (see figure~\ref{fig3c}).
	The peaks of both high- and low-speed streaks could be regarded as extreme $\langle u'\rangle$-events.
	The negative $\partial\langle u'\rangle/\partial x$ of low-speed streaks indicates $\langle u'\rangle$ is approaching its extreme negative events, and vice versa for the positive one.
	It seems to be completely opposite to the high-speed streak case, but they could be uniformly explained by considering $|\langle u'\rangle|$ that the positive $\partial|\langle u'\rangle|/\partial x$ indicates $\langle u'\rangle$ is approaching its extreme events, regardless of its sign.
	Based on this, figure~\ref{fig8b} shows the colored contour surface of $\partial|\langle u'\rangle|/\partial x$ field.
	Compared with $T_{u'\to\tau_\mathrm{w}}$ in figure~\ref{fig3c}, $\partial|\langle u'\rangle|/\partial x$ field presents a remarkable resemblance in spatial distributions, and we therefore have confidence to conclude that the trend of $u'$ (or say, the streak) approaching or receding from its extreme events causally stimulates or suppresses the extreme $\tau_\mathrm{w}$-events.
	It should be emphasized that it is not the extreme $\langle u'\rangle$ events but the processes towards and away from them that exert conclusive influence on the extreme $\tau_\mathrm{w}$-events.
	
	As for $\langle v'\rangle$ and $\langle w'\rangle$, figure~\ref{fig8c} and~\ref{fig8d} show their coupled vector fields in the cross-stream plane $\Delta x=0$.
	The vector fields clearly recognize a pair of counter-rotating streamwise rolls whose cores are exactly located at the regions where both $v'$- and $w'$-causal structures are rather close to zero.
	Despite certain concurrent features, it is indisputable that the conditional streamwise rolls could not explain all observed features of causal structures, particularly regarding the sign discordance between them.
	Following the above analyses of $\langle u'\rangle$, contours of $\partial|\langle v'\rangle|/\partial x$ and $\partial|\langle w'\rangle|/\partial x$ are plotted in figure~\ref{fig8e} and~\ref{fig8f}.
	It is thrilling to find that these two streamwise gradient fields also stay closely aligned with their corresponding LKIF fields (see figure~\ref{fig4b} and~\ref{fig5b}).
	All these findings reveal that the streamwise developments of all velocity components exert influence on the TSD generation in an identical manner.
	Despite the counter-rotating feature of the streamwise rolls pair, they show completely symmetric $\partial|\langle v'\rangle|/\partial x$ and $\partial|\langle w'\rangle|/\partial x$ with respect to the spanwise direction (not shown here for brevity), and consequently indicate their identical causal mechanism on the TSD generation.
	
	\subsection{Causal structures of quadrant-decomposed velocities}\label{sec4.4}
	
	\begin{figure}
		\centering
		\subfigure{
			\begin{overpic}[width=2.3in]{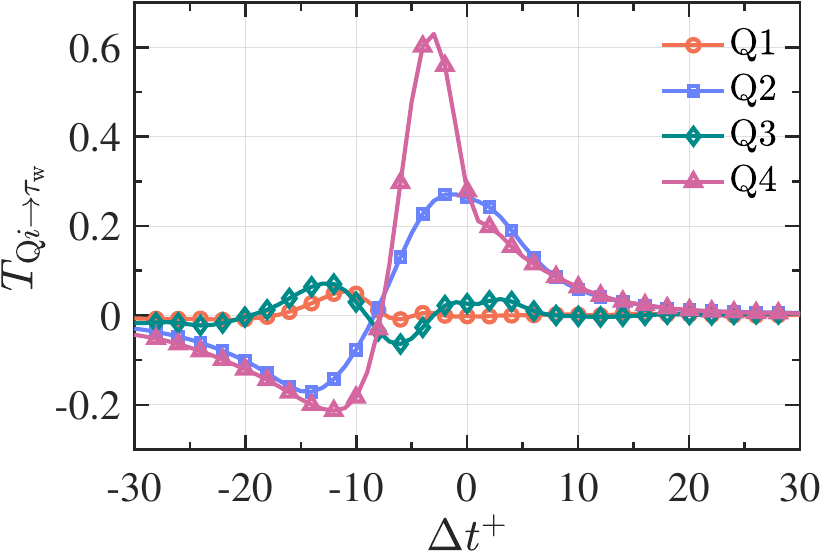}
				\put(1,68){($a$)}
			\end{overpic}\label{fig9a}}
		\subfigure{
			\adjustbox{valign=t, raise=4.48cm}{
			\begin{overpic}[width=2.46in]{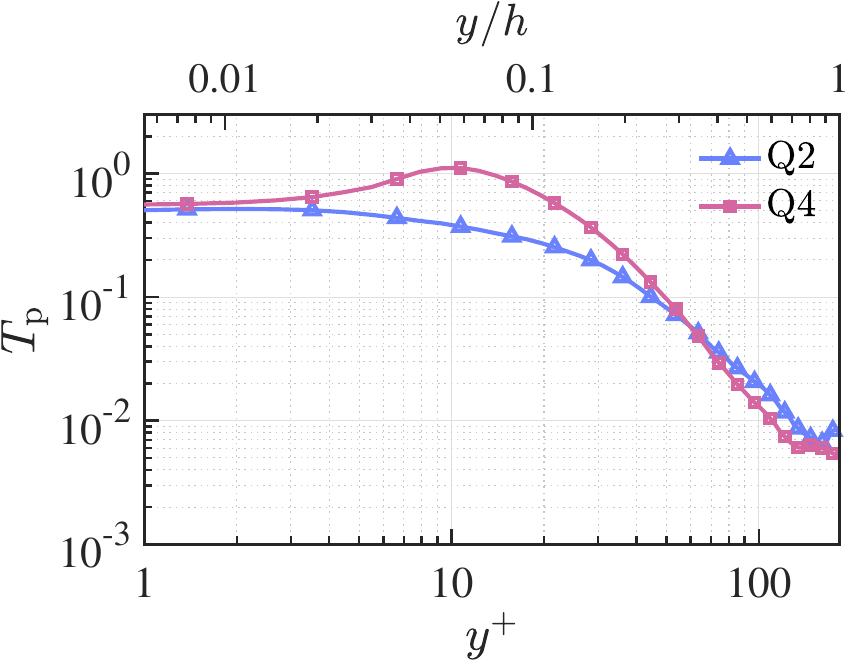}
				\put(3,66.2){($b$)}
			\end{overpic}\label{fig9b}}}
		\caption{LKIF from quadrant motions of $u'v'$ to the target $\tau_\mathrm{w}$.
			($a$) Profiles of $T_{\mathrm{Q}i\to\tau_\mathrm{w}}$ ($i=1\sim4$) versus $\Delta t^+$.
			($b$) Profiles of the positive peak value ($T_{\rm p}$) of both Q2 and Q4 events versus $y$.}
		\label{fig9}
	\end{figure}
	
	The turbulence production in the near-wall region is associated with the bursting process~\citep{Kim1971}, which is also highly related to the TSD generation~\citep{Jeong1997,Flores2010}.
	To investigate the importance of the bursting process on the TSD generation, we decompose $u'v'$ into quadrant motions, including outward interactions Q1 ($u'>0, v'>0$), ejections Q2 ($u'<0, v'>0$), inward interactions Q3 ($u'<0, v'<0$), and sweeps Q4 ($u'>0, v'<0$).
	Using the same computational settings in figure~\ref{fig2}, we calculate LKIF of the quadrant motions ($T_{\mathrm{Q}i\to\tau_\mathrm{w}}$, where $i=1\sim4$) and plot their variations as a function of $\Delta t^+$ in figure~\ref{fig9a}.
	It can be seen that the ejections and sweeps have larger peak values of LKIF than the outward and inward interactions, indicating their predominant contributions to the TSD generation.
	Moreover, the sweeps exert more influences on the TSD generation than the ejections, with the highest peak value among the quadrant motions especially in the near-wall region, as illustrated in figure~\ref{fig9b}.
	This is reasonable as the sweep motions carry massive momentum and energy towards the wall surface, and consequently stimulate the dominant TSD generation.
	The phenomenon corresponds well with the observation in figure~\ref{fig8a} and~\ref{fig8c} that the conditional $\langle\bm{u}'\rangle$ fields show clear sweep events in the vicinity of the target $\tau_\mathrm{w}$.
	Additionally, it is highly consistent with the predominance of sweeps in amplifying velocity fluctuations~\citep{Osawa2024} and in generating high TSD~\citep{Xu2013,Cheng2020}.
	
	\section {Concluding remarks}\label{sec5}
	
	In this study, a novel causal inference method called Liang-Kleeman information flow (LKIF) is first utilized to identify the velocity-induced causal structures in a turbulent channel flow at $Re_\tau\approx183$.
	According to the velocity component, the causal structures are classified into $u'$-, $v'$- and $w'$-causal structures.
	Results suggest that the causal structures identified by LKIF reveal the well-organized spatial-temporal characteristics of flow motions responsible for the generation of turbulent skin-friction drag (TSD), including wall-attachment behaviors, length scales, inclination angles, positive-negative regularities, propagation speeds, and the importance of quadrant motions.
	The comparisons with correlation analysis support the statement that `correlation does not imply causation'.
	More importantly, the signs of the causal and correlation structures have different physical interpretations.
	The positive causal structures serve to promote the growth of information entropy contained in the $\tau_\mathrm{w}$-events, exciting stronger fluctuations of $\tau_\mathrm{w}$ and generating more extreme $\tau_\mathrm{w}$-events.
	By contrast, the negative causal structures have actions to reduce the information entropy, and consequently suppress the formation of extreme $\tau_\mathrm{w}$-events.
	On the other hand, the signs of correlations just reflect the synchrony between the velocity components and the targeted $\tau_\mathrm{w}$.
	Using the velocity fields conditionally averaged with extreme $\tau_\mathrm{w}$-events, we find that their streamwise gradients collapse surprisingly well with the distributions of causal structures, which demonstrates that the trend of streamwise streaks and rolls approaching or receding from their extreme events is the essential factor to promote or suppress the extreme $\tau_\mathrm{w}$-events.
	
	The achievements in this study may help to develop novel drag reduction strategies that precisely control the causal structures within well-defined spatial domains and temporal windows.
	Looking ahead, several directions deserve further studies: 
	(\romannumeral1) more attention needs to be paid to the Reynolds-number effects of the causal structures, aiming to reveal the contributions of large-scale motions on the TSD generation;
	and (\romannumeral2) the causal inference method in use can also be applied for compressible turbulent flows, in order to clarify the generation mechanism of the heat fluxes on the wall surfaces.
	
	\backsection[Acknowledgements]{}
	The funding support of the National Natural Science Foundation of China (under the Grant Nos. 12372221 and 123B2030) is acknowledged.
	Special thanks to Dr. Cheng Cheng at The Hong Kong University of Science and Technology for his utmost assistance.
	We also thank Ph.D. Candidate Carlos Martinez-Lopez at Universidad Polit\'{e}cnica de Madrid for DNS data generation and instructive discussions.
	
	\backsection[Declaration of Interests]{The authors report no conflict of interest.}
	
	\backsection[Supplementary movies]{Movie 1$\sim$5}

	\appendix
	
	\section{Wald test for $T_{2\to1}$}\label{appA}
	The significance of $T_{2\to1}$ derived from Equation~\eqref{eqB5} can be assessed using either a non-parametric method like the block bootstrap or a parametric approach like the Wald test. 
	We favor the Wald test to avoid the tuning parameter selection inherent to the bootstrap.
	A critical step in the Wald test is computing the Fisher information matrix.
	For brevity, we inherit the derivation process and notations in~\cite{Liang2014} until equation (10).
	We note that the notations are only valid in this section and are independent of those in the text (unless explicitly stated).
	Based on the assumptions of: (\romannumeral1) linear model, (\romannumeral2) initial bivariate normal distribution and (\romannumeral3) constant coefficient vectors and matrices, the expression~\eqref{eqB4} is simplified as,
	\begin{equation}\label{eqC1}
		T_{2\to1}=a_{12}\dfrac{C_{12}}{C_{11}},
	\end{equation}
	where $a_{12}$ is estimated via MLE, and therefore $T_{2\to1}$ is estimated as $\hat{T}_{2\to1}=\hat{a}_{12}C_{12}/C_{11}$.
	According to the asymptotic normality of MLE, when $N$ is large, $\hat{a}_{12}$ approximately follows a normal distribution $\mathcal{N}\left(a_{12,\mathrm{true}},\hat{\sigma}_{a_{12}}^2\right)$.
	We can calculate $\hat{\sigma}_{a_{12}}^2$ through Fisher information matrix.
	Since parameters $\left(a_{11},a_{12},f_1,b_1\right)$ and $\left(a_{21},a_{22},f_2,b_2\right)$ are decoupled, we set $\bm{\theta}=\left(a_{11},a_{12},f_1,b_1\right)^{\mathrm T}$ and compute the Fisher information matrix
	\begin{equation*}
		\bm{I}=-\frac{1}{N}\sum_{n=1}^{N}\left.\frac{\partial^2\log\rho_n}{\partial\bm{\theta}\partial\bm{\theta}^{\mathrm T}}\right|_{\bm{\theta}=\bm{\hat{\theta}}},
	\end{equation*}
	where
	\begin{equation*}
		\rho_n=\rho\left(\bm{X}_{n+1}|\bm{X}_n;\bm{\theta}\right)=-\log\left(2\pi\Delta t\right)-\dfrac{1}{2}\log\left(b_1^2b_2^2\right)-\dfrac{\Delta t}{2}\left(\dfrac{R_{1,n}^2}{b_1^2}+\dfrac{R_{2,n}^2}{b_2^2}\right),
	\end{equation*}
	\begin{equation*}
		R_{i,n}=\dot{X}_{i,n}-\left(f_i+a_{i1}X_{1,n}+a_{i2}X_{2,n}\right),
	\end{equation*}
	and $\bm{\hat{\theta}}$ is the optimal parameter vector of MLE.
	Then, we could obtain
	\begin{equation}\label{eqC2}
		\bm{I}=
		\begin{pmatrix}
			\dfrac{\Delta t}{\hat{b}_1^2}\overline{X_1^2} & \dfrac{\Delta t}{\hat{b}_1^2}\overline{X_1X_2} & \dfrac{\Delta t}{\hat{b}_1^2}\overline{X_1} & \dfrac{2\Delta t}{\hat{b}_1^3}\overline{\hat{R}_1X_1}\\
			\dfrac{\Delta t}{\hat{b}_1^2}\overline{X_2X_1} & \dfrac{\Delta t}{\hat{b}_1^2}\overline{X_2^2} & \dfrac{\Delta t}{\hat{b}_1^2}\overline{X_2} & \dfrac{2\Delta t}{\hat{b}_1^3}\overline{\hat{R}_1X_2}\\
			\dfrac{\Delta t}{\hat{b}_1^2}\overline{X_1} & \dfrac{\Delta t}{\hat{b}_1^2}\overline{X_2} & \dfrac{\Delta t}{\hat{b}_1^2} & \dfrac{2\Delta t}{\hat{b}_1^3}\overline{\hat{R}_1}\\
			\dfrac{2\Delta t}{\hat{b}_1^3}\overline{\hat{R}_1X_1} & \dfrac{2\Delta t}{\hat{b}_1^3}\overline{\hat{R}_1X_2} & \dfrac{2\Delta t}{\hat{b}_1^3}\overline{\hat{R}_1} & \dfrac{3\Delta t}{\hat{b}_1^4}\overline{\hat{R}_1^2}-\dfrac{1}{\hat{b}_1^2}
		\end{pmatrix}.
	\end{equation}
	Meanwhile, the MLE conditions lead to $\sum_{n=1}^{N}\partial\log\rho_n/\partial\bm{\theta}|_{\bm{\theta}=\bm{\hat{\theta}}}=\bm0$, i.e.~$\overline{\hat{R}_1X_1}=\overline{\hat{R}_1X_2}=\overline{\hat{R}_1}=0$ and $\hat{b}_1^2=\Delta t\overline{\hat{R}_1^2}$, where $\hat{R}_1=R_1|_{\bm{\theta}=\bm{\hat{\theta}}}$.
	Substitute these MLE conditions into~\eqref{eqC2}, and we can simplify the Fisher information matrix as,
	\begin{equation}\label{eqC3}
		\bm{I}=\dfrac{\Delta t}{\hat{b}_1^2}
		\begin{pmatrix}
			\overline{X_1^2} & \overline{X_1X_2} & \overline{X_1} & 0\\
			\overline{X_2X_1} & \overline{X_2^2} & \overline{X_2} & 0\\
			\overline{X_1} & \overline{X_2} & 1 & 0\\
			0 & 0 & 0 & 2/\Delta t
		\end{pmatrix}.
	\end{equation}
	
	Denote $T=N\Delta t$ as the total duration of series.
	The covariance matrix of $\bm{\hat{\theta}}$ is $\left(N\bm{I}\right)^{-1}$, of which the (2,2) element is exactly
	\begin{equation}\label{eqC4}
		\hat{\sigma}_{a_{12}}^2=\dfrac{\hat{b}_1^2}{T}\dfrac{C_{11}}{C_{11}C_{22}-C_{12}^2}.
	\end{equation}
	Furthermore, due to $\hat{T}_{2\to1}=\hat{a}_{12}C_{12}/C_{11}$, we can obtain the variance of $\hat{T}_{2\to1}$, i.e.~$\hat{\sigma}_{T_{2\to1}}^2=\hat{\sigma}_{a_{12}}^2C_{12}^2/C_{11}^2$.
	Therefore, when $N$ is large, given a significance level $\alpha$, we can accordingly find the confidence interval $\hat{T}_{2\to1}\pm z_{\alpha/2}\hat{\sigma}_{T_{2\to1}}$, where $z_{\alpha/2}$ is the corresponding two-tailed $\alpha$-quantile.
	If zero is not included in the interval, we conclude that $\hat{T}_{2\to1}$ is significant at the $\alpha$ level; otherwise, it is not significant and is set to 0.
	
	\section{Details of LKIF computational framework}\label{appB}
	The computational domain is divided into 105 subdomains (7 along the streamwise and 15 along the spanwise direction) each with a confined size of $L_x'\times L_z'\approx\pi\times\pi/4$ and a grid resolution $N_x'\times N_z'=63\times33$, to compute the statistics and to estimate the LKIF.
	The overlap ratio between two adjacent subdomains is around 50\% in each direction.
	Being aware of the positive wall-normal gradient of the mean velocity, it is expected that the region of influence of turbulent structures grows with the wall-normal height.
	The subdomain size is defined accordingly to ensure that the influence from the upper limit of the studied region is correctly captured for the present Reynolds number.
	After being computed in each subdomain, $T_{\phi\to\tau_\mathrm{w}}$ is averaged over the entire 105 subdomains to obtain a smoothed statistical result.
	All time series share equal length $N_t=8000$, i.e.~the total number of collected snapshots.
	
	\section{Robustness tests of LKIF}\label{appC}
\begin{figure}
		\centering
		\includegraphics[width=2.5in]{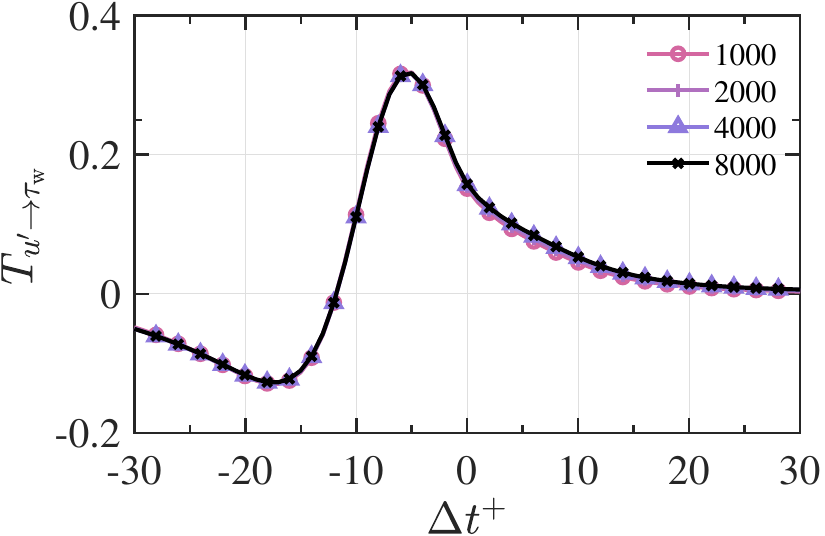}
		\caption{Robustness test of $T_{u'\to\tau_\mathrm{w}}$ with respect to the time series length (from 1000 to 8000). The example $T_{u'\to\tau_\mathrm{w}}$ is at $(\Delta x,y^+,\Delta z)=(0,20,0)$.}
		\label{figS1}
	\end{figure}

	Robustness tests of LKIF with respect to the time series length are performed.
	Four cases of $T_{u'\to\tau_\mathrm{w}}$ are computed, with the length of $u'$ and $\tau_\mathrm{w}$ series being 1000, 2000, 4000 and 8000, respectively.
	For brevity, we only present the result of $T_{u'\to\tau_\mathrm{w}}$ at $(\Delta x,y^+,\Delta z)=(0,20,0)$ where the fourth line in figure~\ref{fig2a} locates.
	As shown in figure~\ref{figS1}, four curves collapse pretty well at any time lag, suggesting that the LKIF computation is robust when the series length is larger than 1000 (i.e. the time $T^+$ is longer than 1000).   
	
	\section{Significance tests of causal structures}\label{appD}
	
\begin{figure}
		\centering
		\includegraphics[width=4in]{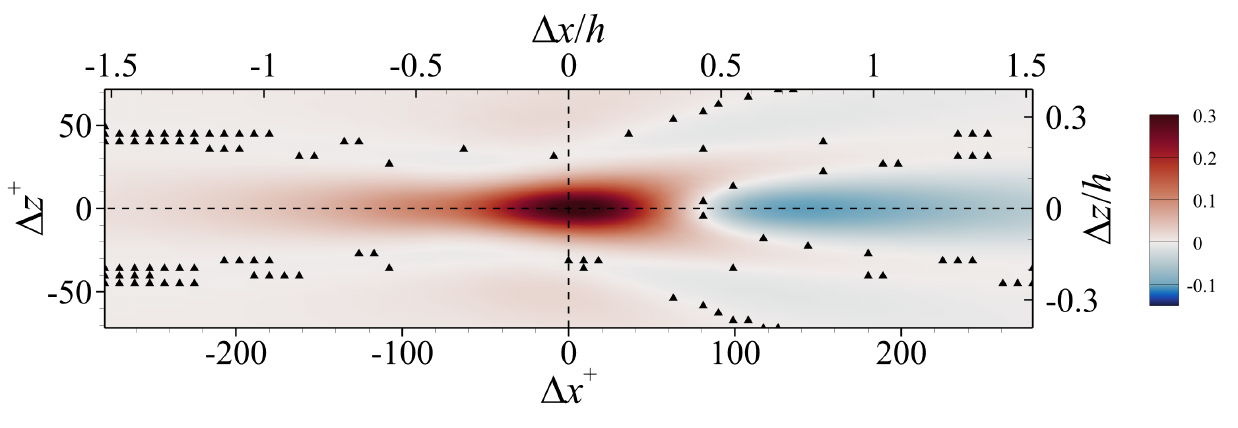}
		\caption{Colored contour of $T_{u'\to\tau_\mathrm{w}}$ at $y^+=20$ at the time lag $\Delta t^+=-5$.
			Triangles indicate insignificant points where $T_{u'\to\tau_\mathrm{w}}$ should be set to zero.}
		\label{figS2}
	\end{figure}

	Significance tests have been performed for all computed LKIFs in the text.
	For brevity, we only present the result of $T_{u'\to\tau_\mathrm{w}}$ at the plane $y^+=20$.
	Using a significance level $\alpha=0.01$, the corresponding confidence interval is given by $T_{u'\to\tau_\mathrm{w}}\pm2.576\hat{\sigma}_{T_{u'\to\tau_\mathrm{w}}}$, where $\hat{\sigma}_{T_{u'\to\tau_\mathrm{w}}}$ is determined by Equation~\eqref{eqC4}.
	Recall that if zero is included in the confidence interval, $T_{u'\to\tau_\mathrm{w}}$ is insignificant and is consequently set to zero.
	Figure~\ref{figS2} illustrates the distribution of insignificant points of $T_{u'\to\tau_\mathrm{w}}$ at $y^+=20$ and at the time lag $\Delta t^+=-5$.
	These points are exclusively located along the curves where $T_{u'\to\tau_\mathrm{w}}$ values are rather close to zero.
	The difference value between the original and zero-modified $T_{u'\to\tau_\mathrm{w}}$ is exactly the value of insignificant $T_{u'\to\tau_\mathrm{w}}$.
	The maximal observed difference merely accounts for about 1\% of the maximal $T_{u'\to\tau_\mathrm{w}}$ magnitude at $y^+=20$.
	Therefore, the zero-modification procedure has little impact on the continuity of original $T_{u'\to\tau_\mathrm{w}}$ field.
	We can declare with 99\% confidence that causal structures identified by $T_{u'\to\tau_\mathrm{w}}$ are significant.
	Results of significance tests for $T_{u'\to\tau_\mathrm{w}}$ at other wall-normal positions, as well as for LKIFs from other flow quantities, exhibit similar characteristics to those presented above.
	
	\section{Causal structures at a higher friction Reynolds number}\label{appE}
	
	We adopt another DNS database of a turbulent channel flow at $Re_\tau\approx548$ to compute LKIF in a same manner as how we do in the main text.
	The size of the computational domain is $L_x\times L_z\times L_y=4\pi h\times2\pi h\times2h$ (where $h=1$ is the channel half-height), with a grid resolution of $N_x\times N_z\times N_y=768\times512\times384$ in the streamwise, spanwise and wall-normal directions.
	The flow fields are stored with a time interval of $\Delta t_\mathrm{s}^+=0.9015$.
	
\begin{figure}
		\centering
		\includegraphics[width=2.5in]{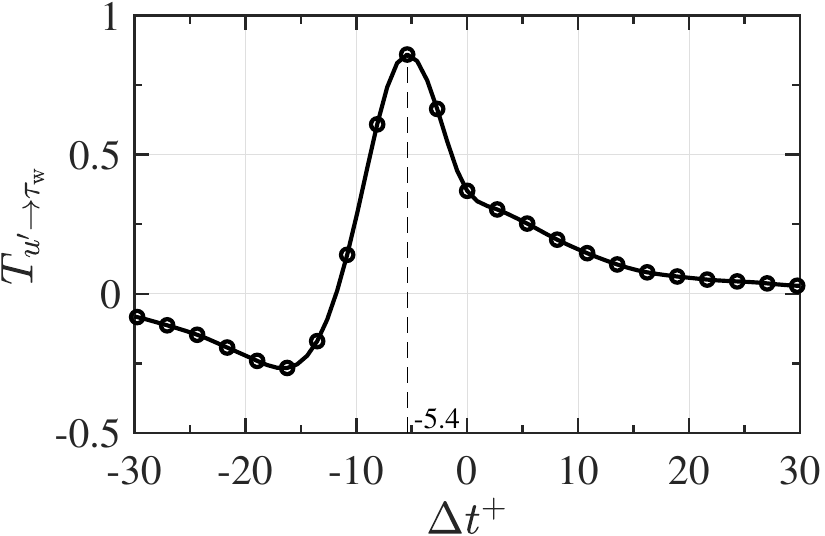}
		\caption{Profiles of $T_{u'\to\tau_\mathrm{w}}$ from $u'$ at $y^+=20$ (right above the target $\tau_\mathrm{w}$) to the target $\tau_\mathrm{w}$ versus $\Delta t^+$, based on the database at $Re_\tau\approx548$.}
		\label{figS3}
	\end{figure}

\begin{figure}
		\centering
		\subfigure{
			\begin{overpic}[width=3.5in]{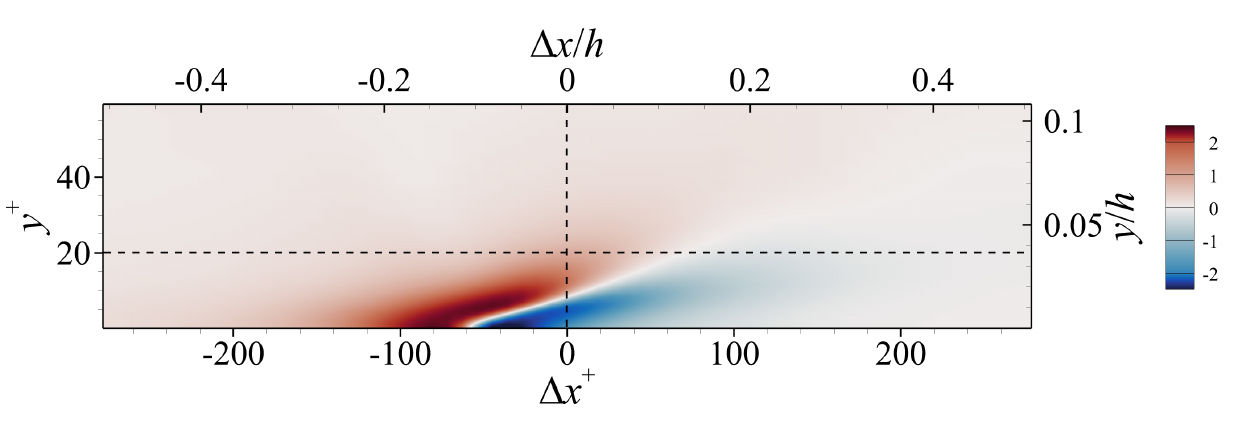}
				\put(1,26){($a$)}
			\end{overpic}\label{figS4a}}
		\subfigure{
			\begin{overpic}[width=3.5in]{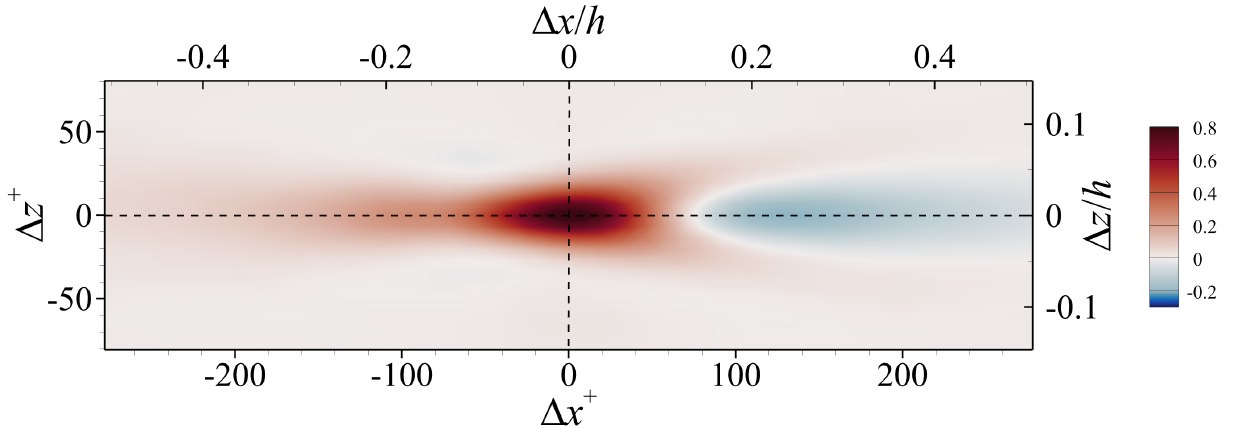}
				\put(1,28){($b$)}
			\end{overpic}\label{figS4b}}
		\caption{Spatial distribution of $T_{u'\to\tau_\mathrm{w}}$ at the time lag $\Delta t^+=-5.4$, based on the database at $Re_\tau\approx548$.
			($a$) Colored contour of $T_{u'\to\tau_\mathrm{w}}$ at the plane $\Delta z=0$.
			($b$) Colored contour of $T_{u'\to\tau_\mathrm{w}}$ at the plane $y^+=20$.}
		\label{figS4}
	\end{figure}

\begin{figure}
		\centering
		\subfigure{
			\begin{overpic}[width=3.5in]{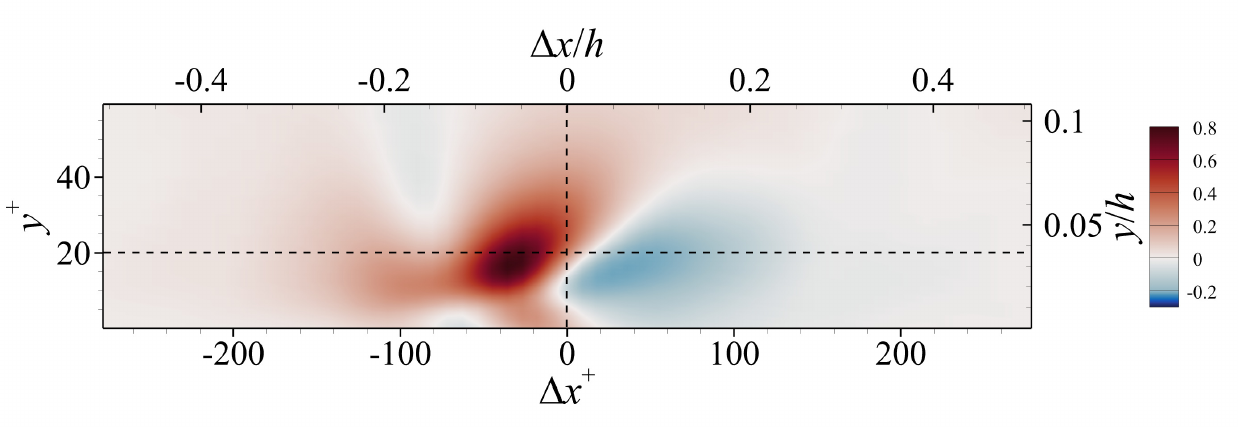}
				\put(1,26){($a$)}
			\end{overpic}\label{figS5a}}
		\subfigure{
			\begin{overpic}[width=3.5in]{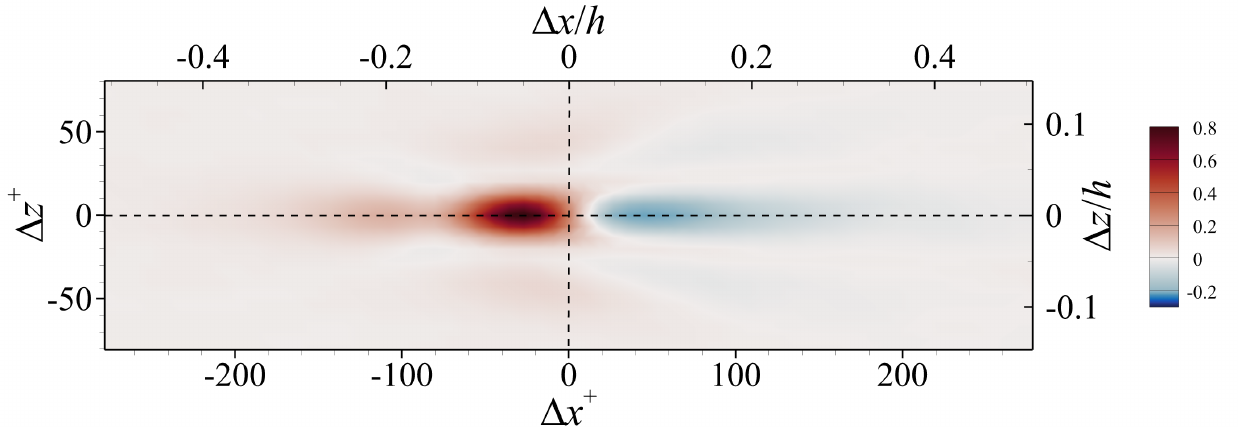}
				\put(1,28){($b$)}
			\end{overpic}\label{figS5b}}
		\caption{Spatial distribution of $T_{v'\to\tau_\mathrm{w}}$ at the time lag $\Delta t^+=-5.4$, based on the database at $Re_\tau\approx548$.
			($a$) Colored contour of $T_{v'\to\tau_\mathrm{w}}$ at the plane $\Delta z=0$.
			($b$) Colored contour of $T_{v'\to\tau_\mathrm{w}}$ at the plane $y^+=20$.}
		\label{figS5}
	\end{figure}

\begin{figure}
		\centering
		\subfigure{
			\begin{overpic}[width=3.5in]{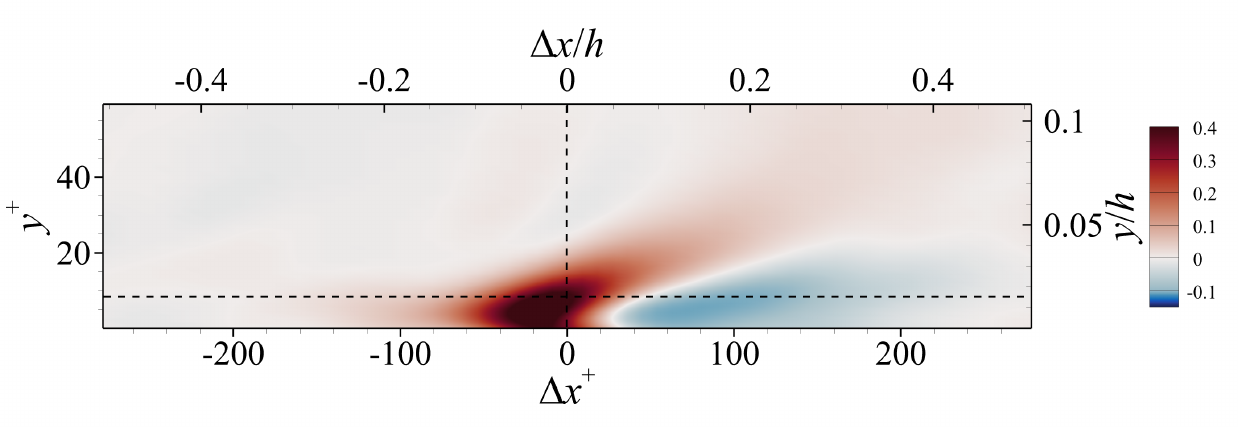}
				\put(1,26){($a$)}
			\end{overpic}\label{figS6a}}
		\subfigure{
			\begin{overpic}[width=3.5in]{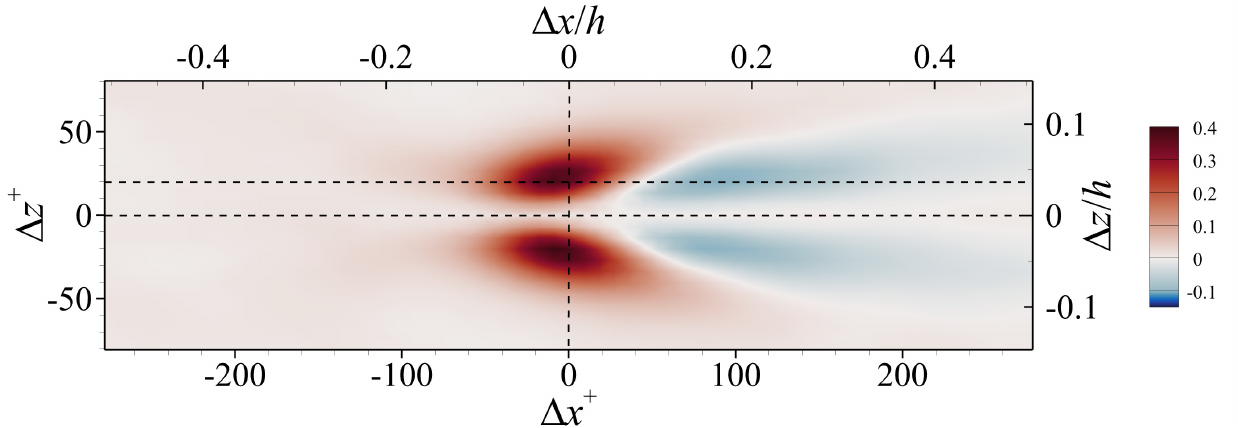}
				\put(1,28){($b$)}
			\end{overpic}\label{figS6b}}
		\caption{Spatial distribution of $T_{w'\to\tau_\mathrm{w}}$ at the time lag $\Delta t^+=-5.4$, based on the database at $Re_\tau\approx548$.
			($a$) Colored contour of $T_{w'\to\tau_\mathrm{w}}$ at the plane $\Delta z^+=20$.
			The horizontal dashed line denotes $y^+=8$.
			($b$) Colored contour of $T_{w'\to\tau_\mathrm{w}}$ at the plane $y^+=8$.
			The upper horizontal dashed line denotes $\Delta z^+=20$.}
		\label{figS6}
	\end{figure}

	Figure~\ref{figS3} shows the variations of $T_{u'\to\tau_\mathrm{w}}$ as a function of the time lag $\Delta t^+$, where $u'$ locates at $y^+=20$ and just above the target $\tau_\mathrm{w}$.
	The $T_{u'\to\tau_\mathrm{w}}$ reaches its positive peak at $\Delta t^+=-5.4$, which is close to the result shown in figure~\ref{fig2a}.
	Analogous to the main text, the $u'$-, $v'$-, and $w'$-causal structures at specific planes at the time lag $\Delta t^+=-5.4$ are plotted in figures~\ref{figS4},~\ref{figS5} and~\ref{figS6}, respectively.
	All causal structures at $Re_\tau\approx548$ exhibit considerably similar characteristics with those at $Re_\tau\approx183$ shown in the main text.
	
	\bibliographystyle{jfm}
	\bibliography{jfm}
	
\end{document}